\pdfoutput=1 
\documentclass[11pt]{article}
\usepackage{fullpage}
\usepackage[pagebackref, colorlinks = true, linkcolor = blue, urlcolor  = blue, citecolor = purple, bookmarks, hypertexnames=false]{hyperref}
\usepackage[bookmarks]{hyperref}
\usepackage{amssymb}
\usepackage{amsmath}
\usepackage{amsthm}
\usepackage{graphicx,color,colordvi}
\usepackage{bbm}
\usepackage{varioref}
\usepackage[capitalise]{cleveref}
\usepackage{stmaryrd}
\usepackage[utf8]{inputenc}
\usepackage[blocks]{authblk}
\usepackage{dsfont}
\usepackage{mathtools}
\usepackage{wrapfig}
\usepackage[english]{babel}
\usepackage{physics}
\usepackage{braket}
\usepackage[table]{xcolor}
\usepackage[center]{caption}
\usepackage{tikz}
\usetikzlibrary{shapes,arrows,patterns,positioning,shapes.geometric,arrows.meta,decorations.markings}
\usepackage{ifthen}
\usepackage{pgfplots}
\pgfplotsset{compat=1.9}
\usetikzlibrary{shapes,arrows.meta}
\usetikzlibrary{positioning}
\usetikzlibrary{shapes.geometric}
\RequirePackage[framemethod=default]{mdframed}
\usepackage[margin=1in]{geometry}
\usepackage{comment}
\usepackage{url}
\usepackage{ upgreek }
\usepackage{makecell}
\usepackage{fancyref} 
\usepackage{float}
\usepackage{enumitem}
\usepackage{autonum}
\usepackage{subcaption}
\newfloat{algorithm}{t}{lop}
\usepackage{array,rotating ,makecell, multirow, tabularx}
\usepackage{scrextend}
\usepackage{mathrsfs}
\usepackage{enumitem}
\usepackage{soul}

\usepackage{IEEEtrantools}
\usepackage[normalem]{ulem}
\usepackage{cancel}

\newcommand{\boxElement}[7]{
    \draw[#5, rounded corners=3pt, thick] (#1,#2) rectangle (#1 + #3,#2 + #4);
    \fill[#5, rounded corners=3pt, opacity=0.1] (#1,#2) rectangle (#1 + #3,#2 + #4);
    
    \node[fill=white, rounded corners=3pt, anchor=south west, inner sep=2pt] at (#1 + 0.25,#2 + #4 - 0.25) {\textcolor{#5}{\textbf{#6}}};

    \node[anchor=north west, text width={#3cm - 0.5cm}, align=left] at (#1 + 0.25, #2 + #4 - 0.25) {#7\hfill~};   }
\tikzset{plus arrow/.style={
        postaction={decorate},
        decoration={
            markings,
            mark=at position 1 with {
                \node[draw, circle, thick, fill=white, inner sep=0pt, minimum size=8pt] {\textbf{+}};}}}}

\DeclareMathOperator{\Pos}{Pos}

\DeclareMathOperator{\id}{id}

\newcommand{\1}{\ensuremath{\mathbbm{1}}}

\usepackage[capitalise]{cleveref}
\newtheoremstyle{newdefinition}{}{}{\normalfont}{}{\bfseries}{}
{ }
{\thmname{#1} \thmnumber{#2}\thmnote{ (#3)}}

\newtheoremstyle{newplain}{}{}{\itshape}{}{\bfseries}{}{1em}
{\thmname{#1} \thmnumber{#2}\thmnote{ (#3)}}

\newtheoremstyle{newremark}{}{}{\normalfont}{}{\bfseries}{}{1em}
{\thmname{#1}}

\theoremstyle{newdefinition}

\theoremstyle{newplain}
\newtheorem{theorem}{Theorem}[section]  
\newtheorem{lemma}[theorem]{Lemma}
\newtheorem{proposition}[theorem]{Proposition}
\newtheorem{corollary}[theorem]{Corollary}

\theoremstyle{newdefinition}
\newtheorem{definition}[theorem]{Definition}
\newtheorem{notation}{Notation}
\newtheorem{remark}[theorem]{Remark}

\crefname{theorem}{Theorem}{Theorems}
\crefname{lemma}{Lemma}{Lemmas}
\crefname{proposition}{Proposition}{Propositions}
\crefname{corollary}{Corollary}{Corollaries}
\crefname{remark}{Remark}{Remarks}
\crefname{example}{Example}{Examples}
\crefname{conjecture}{Conjecture}{Conjectures}
\crefname{definition}{Definition}{Definitions}
\crefname{notation}{Notation}{Notations}
\usepackage{autonum}   

\newtheoremstyle{myplain}{5pt}{5pt}{\itshape}{0pt}{\bfseries}{}{5pt plus 1pt minus 1pt}{}
\theoremstyle{myplain}
\newtheorem*{theorem*}{Theorem}
\newtheorem*{corollary*}{Corollary}

\DeclareMathOperator{\cH}{\mathcal{H}}

\DeclareMathOperator{\cS}{\mathcal{S}}

\DeclareMathOperator{\cM}{\mathcal{M}}

\DeclareMathOperator{\cX}{\mathcal{X}}
\DeclareMathOperator{\cY}{\mathcal{Y}}

\DeclareMathOperator{\cD}{\mathcal{D}}
\DeclareMathOperator{\cN}{\mathcal{N}}

\title{Additivity and chain rules for quantum entropies \\
 via multi-index Schatten norms}

\date{\today}

\author[1]{Omar Fawzi\thanks{Email: omar.fawzi@ens-lyon.fr}}
\author[2]{Jan Kochanowski\thanks{Email: jan.kochanowski@inria.fr}}
\author[2]{Cambyse Rouzé\thanks{Email: rouzecambyse@gmail.com}}
\author[3]{Thomas Van Himbeeck\thanks{Email: thomas.van-himbeeck@inria.fr}}

\affil[1]{Inria, ENS Lyon, UCBL, LIP, F-69342 Lyon Cedex 07, France}
\affil[2]{Inria, Télécom Paris - LTCI, Institut Polytechnique de Paris, 91120 Palaiseau, France}
\affil[3]{Inria Paris, France}

\date{\today}

\begin{document}

\maketitle

\begin{abstract}
The primary entropic measures for quantum states are additive under the tensor product. In the analysis of quantum information processing tasks, the minimum entropy of a set of states, e.g., the minimum output entropy of a channel, often plays a crucial role. A fundamental question in quantum information and cryptography is whether the minimum output entropy remains additive under the tensor product of channels. Here, we establish a general additivity statement for the optimized sandwiched R\'enyi entropy of quantum channels. For that, we generalize the results of [Devetak, Junge, King, Ruskai, CMP 2006] to multi-index Schatten norms. As an application, we strengthen the additivity statement of [Van Himbeeck and Brown, 2025] thus allowing the analysis of time-adaptive quantum cryptographic protocols. In addition, we establish chain rules for R\'enyi conditional entropies that are similar to the ones used for the generalized entropy accumulation theorem of [Metger, Fawzi, Sutter, Renner, CMP 2024].

\end{abstract}

\newpage 
\setcounter{tocdepth}{2}        
\renewcommand{\contentsname}{Table of Contents}
\tableofcontents 

\newpage

\section{Introduction}

Entropy is a cornerstone of information theory, governing fundamental limits in communication, compression and statistical inference. 
Entropic quantities often behave extensively when evaluated on composite systems, a property encapsulated by chain rules, additivity or uncertainty relations. Yet, establishing such statements often presents significant challenges, a fact that is particularly true in the quantum setting. A powerful perspective emerges by recognizing that entropies naturally arise as logarithms of certain $L_p$-norm quantities, allowing deep functional analytic methods to be used in information theory, for example to establish \emph{entropy power} inequalities or \emph{uncertainty principles}—closely tied to the behavior of $L_p$ norms under convolution or Fourier transform
\cite{Lieb.1978,Dembo.1991}.

Of particular importance to cryptography is the conditional $\alpha$-R\'{e}nyi entropy of a bipartite distribution $p_{AB}$, $\alpha\ge 1$ \cite{Arimoto.1977}: 
\begin{align}
H^{\uparrow}_\alpha(A|B)_p:=\frac{\alpha}{1-\alpha}\log \left[\sum_b p(b)\, \left(\sum_ap(a|b)^\alpha\right)^{\frac{1}{\alpha}}\right]\,.
\end{align}

In the above, the quantity inside the logarithm can be interpreted as the $\ell_{(1,\alpha)}$-norm of the function $(a,b)\mapsto p(a,b)$. Similarly, multipartite extensions of conditional R\'{e}nyi entropies would involve multi-index $\ell_p$ norms: given a vector $v\in\bigotimes_{i=1}^k\mathbb{C}^{d_i}$ and indices $\{p_i\}_{i=1}^n$ the $\ell_{(p_1,...,p_k)}$ norm of $v$ is defined as 
\begin{align} 
    \|v\|_{(p_1,...,p_k)} := \left(\sum_{a_1=1}^{d_1}\left(\sum_{a_2=1}^{d_2}\,... \left(\sum_{a_k=1}^{d_k}\big|v_{a_1a_2...a_k}\big|^{p_k}\right)^{p_{k-1}/p_k} ...\,\right)^{p_1/p_2}\right)^{{1/p_1}}\,.
\end{align}

While this correspondence is straightforward in the classical setting, it becomes more subtle in the quantum case due to the absence of a preferred basis. Note that in the case of a single index, i.e., $k=1$, the Schatten norm $\cS_p(\mathbb{C}^d)$ defined by $\| X \|_{p} = \tr[|X|^p]^{1/p}$ is a natural non-commutative extension of the norm $\ell_p$. However, the multi-index, i.e., $k \geq 2$ case, is more difficult as direct extensions such as  
\[
 (\tr_1([\tr_2[...(\tr_k[|X|^{p_k}])^{{p_{k-1}/p_k}}...])^{{p_1/p_2}}])^{{1/p_1}}
\]
for operators \(X\) acting on \( \bigotimes_{i}\mathbb{C}^{d_i}\), do not define norms \cite{Devetak.2006}. As an attempt to restore the norm property via operator space theoretic methods, Pisier introduced the concept of \textit{operator-valued Schatten norms} in his seminal work \cite{Book.Pisier.1998}. 
An operator space $\cX$ consists of a family of norms $M_{n}(\cX)$ indexed by $n$ on the space of $n \times n$ matrices with entries in $\cX$. Such a family should satisfy natural conditions for an operator norm.
For an arbitrary operator space $\cX$, Pisier proposed the following extension of $p$-Schatten norms to $M_{d_1}\otimes \mathcal{X}$, whose form is reminiscent of H\"{o}lder's inequality: 
\begin{align}
    \|X\|_{\mathcal{S}_{p_1}[\mathbb{C}^{d_1},\mathcal{X}]}= \inf_{\underset{X=FYG}{F,G\in \mathcal{S}_{2p_1}(\mathbb{C}^{d_1}), Y\in M_{d_1}(\cX)}}\|F\|_{2p_1}\|Y\|_{M_{d_1}(\cX)}\|G\|_{2p_1}\,,
\end{align}
Choosing $\mathcal{X}$ itself as a Schatten space $\mathcal{S}_{p_2}(\mathbb{C}^{d_2})$, Pisier's formula provides a means to define a version of $\ell_{(p_1,p_2)}$ for operators (see Theorem \ref{thm:expression-pq} below). By iterating this procedure, one gets a natural quantum extension of $\ell_{(p_1,\dots\,p_k)}$:
\begin{align}
    \|X\|_{(p_1,p_2\cdots\, p_n)} = \|X\|_{\mathcal{S}_{p_1}[ \mathbb{C}^{d_1}, \mathcal{S}_{p_2}[ \mathbb{C}^{d_2} \cdots \, \mathcal{S}_{p_n}(\mathbb{C}^{d_n})]\cdots ]}\,.
\end{align}

A major connection between these multi-index Schatten norms and quantum entropies was established in \cite{Devetak.2006}: the $(1,\alpha)$-Schatten norm can be related to von Neumann entropies by taking appropriate derivatives. Later, the sandwiched R\'enyi conditional entropies were defined as
\begin{align}
    H^\uparrow_\alpha(A|B)_\rho := -\inf_{\sigma_B\in \mathcal{D}(\mathcal{H}_B)}D_{\alpha}(\rho_{AB}\|\1_A\otimes\sigma_B), 
\end{align} 
where $D_\alpha$ corresponds to the sandwiched R\'{e}nyi divergence of order $\alpha\ge 1$ \cite{Lennert.2013,Wilde.2014,Book.Tomamichel.2016} (see Section \ref{renyidefs}). As observed in \cite{Beigi.2023}, such conditional entropy can directly be related to $(1,\alpha)$-Schatten norms:
\begin{align} \label{equ:intro.conditional.entropy}
   H^\uparrow_\alpha(A|B)_\rho = \frac{\alpha}{1-\alpha}\log\|\rho_{BA}\|_{(1,\alpha)}.
\end{align}

\subsection{Main results}

\subsubsection{Multiplicativity of completely bounded norms}

In addition to pointing out the deep connection between Pisier spaces and conditional quantum entropies, in \cite{Devetak.2006}, the authors proved the multiplicativity of the completely bounded norms between $L_p$ spaces for completely positive (CP) maps: in particular, given two channels $\Phi_1:Q_1\to S_1$, $\Phi_2:Q_2\to S_2$ and any $1\leq p,q \le \infty$, \cite{Devetak.2006} showed that the product channel $\Phi=\Phi_1\otimes\Phi_2:Q\to S$, with $Q = Q_1Q_2$ and $S=S_1S_2$ satisfies 
\begin{align}
    \|\Phi\|_{cb, \mathcal{S}_p(\mathcal{H}_{Q}) \to \mathcal{S}_q(\mathcal{H}_{S})} = \|\Phi_1\|_{cb, \mathcal{S}_p(\mathcal{H}_{Q_1}) \to \mathcal{S}_q(\mathcal{H}_{S_1})}\|\Phi_2\|_{cb, \mathcal{S}_p(\mathcal{H}_{Q_2}) \to \mathcal{S}_q(\mathcal{H}_{S_2})}\,,
\end{align}
where $\|\cdot\|_{cb,\mathcal{X}\to\mathcal{Y}}$ denotes the completely bounded norm between two operator spaces.
Generalizing the multiplicativity to arbitrary multi-index Schatten operator spaces, we prove in Theorem~\ref{thm:general.multiplicativity}, for any CP maps $\{\Phi_i:Q_i\to S_i\}$ and numbers $1\leq q_i,p_i\leq\infty$,
\begin{align}
    \bigg\|\bigotimes_{i=1}^n\Phi_i\bigg\|_{cb,(q_1,...,q_n)\to (p_1,...,p_n)} = \prod_{i=1}^n\|\Phi_i\|_{cb,q_i\to p_i}\,.
\end{align}

\subsubsection{Additivity of output $\alpha$-Rényi conditional entropy}

Reinterpreting the multiplicativity result of \cite{Devetak.2006} in terms of Rényi $\alpha$-entropies and choosing $(p,q) = (1,\alpha)$ yields the additivity of the output Rényi $\alpha$-entropy of CP maps:
\begin{align}
\inf_E\inf_{\rho_{EQ}}H_\alpha^\uparrow(S|E)_{\Phi(\rho)}=\inf_E\inf_{\rho_{EQ_1}}H_\alpha^\uparrow(S_1|E_1)_{\Phi_1(\rho)}+\inf_E\inf_{\rho_{EQ_2}}H_\alpha^\uparrow(S_2|E_2)_{\Phi_2(\rho)}\,.
\end{align}
In \cite{Himbeeck.2025} using different techniques, the authors proved that for a CP map $\Phi:Q\to RS$ with classical output registers $R,S$, the $n$-fold tensor product map $\Phi^{\otimes n} : Q^n\to R^nS^n$ satisfies
\begin{align}
\inf_E\inf_{\rho_{EQ^n}}H_\alpha^\uparrow(S^n|R^nE)_{\Phi^{\otimes n}(\rho)}
        &= n \cdot \inf_E\inf_{\rho_{EQ}}H_\alpha^\uparrow(S|RE)_{\Phi(\rho)}\,.
\end{align}
This result was referred to as IID reduction since it implies that the minimizer of the LHS takes the tensor product form $\rho_{E^\prime Q}^{\otimes n}$ with $E = {E^\prime}^n$, representing identically and independently distributed quantum systems. In terms of operator norms, note that this is equivalent to $1\to (1,p)$ norms $\|\Phi^{\otimes n}\|_{cb, 1 \to (1,p)} = \|\Phi\|_{cb, 1 \to (1,p)}^n$.

We generalize both results and show in Theorem~\ref{thm:mainchainrule} that for any CP maps $\Phi_i:Q_i\to R_iS_i$, the product channel $\Phi^n=\bigotimes_{i\le n} \Phi_i :Q^n\to R^n S^n$ with $Q^n = Q_1\cdots Q_n$, $R^n = R_1\cdots R_n$, $S^n=S_1 \cdots S_n$, satisfies 
\begin{align}\label{eq:generalized.IID.intro}
\inf_E\inf_{\rho_{EQ^n}}H_\alpha^\uparrow(S^n|R^nE)_{\Phi^n(\rho)}
        &= \sum_i \inf_E\inf_{\rho_{EQ_i}}H_\alpha^\uparrow(S_i|R_iE)_{\Phi_i(\rho)}\,.
\end{align}
This is equivalent to $\|\Phi^n\|_{cb, 1 \to (1,p)} = \prod_{i=1}^n \|\Phi_i\|_{cb, 1 \to (1,p)}$.
Note here the difference between the product channel $\Phi^{\otimes n}$ made up of $n$ copies of the same map, considered in \cite{Himbeeck.2025} and the more general product of $n$ different channels denotes as $\Phi^n$. We will continue this notation convention throughout this work.

\subsubsection{Chain rule for the $\alpha$-Rényi entropy:}

As a consequence of the chain rule for $\alpha$-R\'{e}nyi entropies derived in \cite[Lemma 3.6]{Metger.2024}, for any two channels $\Phi_1:Q_1\to S_1 $ and $\Phi_2:Q_2\to S_2$ with $\Phi=\Phi_1\otimes\Phi_2$, $\alpha\in(1,2)$ and any state $\rho_{QT}$,
\begin{align}
H_\alpha(TS_2|S_1)_{\Phi(\rho)}\ge H_\alpha(T|Q_1)_\rho+\inf_{\sigma_{Q\tilde{Q}}}H_{\frac{1}{2-\alpha}}(S_2|S_1\tilde{Q})_{\Phi(\sigma)}
\end{align}
for a purifying system $\tilde{Q}$ of $Q=Q_1Q_2$, where we recall that the non-optimized R\'{e}nyi conditional entropy is defined as $H_\alpha(A|B)_\rho:=-D_\alpha(\rho_{AB}\|\1_A\otimes \rho_B)$, and where the infimum on the right-hand side is over all quantum states on $Q\tilde{Q}$.\footnote{The inequality proved in \cite{Metger.2024} holds in fact more generally for channels satisfying a certain non-signalling property.} Exploiting operator valued Schatten spaces, we derive a similar, yet seemingly tighter inequality for the optimized R\'{e}nyi entropy (see \cref{cor:generalized.eat.chain.rule}): for any $\alpha\ge 1$ and any state $\rho_{QT}$,
\begin{align}\label{eq:intro.generalized.EAT}
H^\uparrow_\alpha(TS_2|S_1)_{\Phi(\rho)}\ge H_\alpha^\uparrow(T|Q_1)_\rho+\inf_{\sigma_{Q_1Q_2\tilde{Q}}}H^\uparrow_\alpha(S_2|S_1\tilde{Q})_{\Phi(\sigma)}.
\end{align}

Moreover, 
we derive the following variant for two-output channels (see \cref{cor:Chain.Rules}): For any state $\rho_{QT}$ and any quantum channel $\Phi:Q\to RS$,
\begin{align}
\label{eq:chain.rule.intro}
H^{\uparrow}_\alpha(TS|R)_{\Phi 
(\rho)} - H^{\uparrow}_\alpha(T|Q)_\rho &\geq   \inf_{\sigma_Q} H^{\uparrow}_\alpha(S|R)_{\Phi(\sigma)},
\end{align} 
Once again, this bound can be compared to~\cite[Lemma 3.6]{Metger.2024}: while our result is less general, it is directly stated for the optimized R\'enyi entropy, suffers no loss in the parameter $\alpha$, and does not require optimizing over purifications on its right-hand side.

\subsubsection{Application to quantum cryptography}

Quantum entropies have many applications in quantum cryptography and quantum key distribution \cite{Xu_2020} in particular. Historically, their theoretical study has gone hand in hand with the development of rigourous and efficient security proof for QKD \cite{Renner.2006, Arnon.2018,
Metger.2023,Himbeeck.2025}. In the present article, we follow this trend and show (\cref{thm:time_adaptive_protocol}) how our new additivity results for output Rényi entropies lead to a new family of security proofs for quantum protocols that are time-dependent and that are subjected to time-dependent experimental conditions. 

In quantum key distribution, the assumption is usually made that the protocol does not vary with time. This setting is well-adapted to static scenarios, such as protocols deployed over optical fibers where fluctuations can be neglected. This is not the case however for free-space implementations such as satellite QKD \cite{Liao_2017} where the trajectory of the satellite and atmospheric turbulence lead to noise patterns that vary with time. 

It is possible to apply a static security proof in cases where the noise on the channel varies but not the protocol itself. While the security statements will not be affected in this case, the performance may not be optimal. Here, we show that in cases where the noise varies with time in a way that is predictable, we can achieve higher secret key rates than with traditional static security proofs. Moreover, our new security proof also applies to protocols that change over times, which opens up the possibility of designing new QKD protocols that are specially tailored to time-dependent scenarios.

\subsection{Structure of the paper}

Each of our main results is stated in two ways, highlighting the correspondence between the functional and entropic settings discussed above: first in terms of submultiplicativity of (completely bounded) operator norms between operator-valued Schatten spaces of CP maps, then via \eqref{equ:intro.conditional.entropy} as inequalities for the output Rényi conditional entropy of quantum channels.

In \cref{sec:Generalized.Variational.Expressions}, we recall the main notions of Pisier's formalism, and in particular \cref{thm:PisierOriginal}, which we extensively use to derive  general variational formulas with the goal to make triple-index Schatten norms more tractable. 
We do this by introducing a systematic way to derive variational formulas for Schatten norms for arbitrary numbers of indices in \cref{thm:VariationalSpOperatorspace}, and later focus on the case of 3 indices in \cref{cor:variationalFormula.2} and \cref{cor:variationalFormula}.

In \cref{sec:Main.Results}, we apply these bounds to derive our main results. First, we derive in \cref{lem:reductionLemma} a generalization of \cite[Lemma 5]{Devetak.2006} to two-output CP maps of the form $\Phi:Q\to RS$. Informally, this result states that ``identities to the right'' do not have an effect on CP maps between operator-valued Schatten spaces.

The chain rule that results from this is stated in \cref{Cor:rightIdentity}, see also \eqref{eq:chain.rule.intro}. 
We then prove a general ordered multiplicativity result for completely bounded norms in \cref{thm:general.multiplicativity}, which follows from the aforementioned \cref{lem:reductionLemma}.
A direct consequence of it is the entropic chain for product maps already stated in \eqref{eq:intro.generalized.EAT}, see \cref{cor:generalized.eat.chain.rule}.
Our last main technical result is a (non-ordered) multiplicativity result for 
$1\to(1,p)$-completely bounded norms in \cref{thm:mainchainrule}. We further show multiplicativity under arbitrary linear input constraints \cref{thm:Restricted.multiplicativity} and for weights to get the additivity statement for the minimum output entropy \cref{thm:IIDreduction}. This is the generalization to tensor products of arbitrary quantum channels of the IID reduction from \cite{Himbeeck.2025} already hinted at in \eqref{eq:generalized.IID.intro}. This result is applied to quantum key distribution in \cref{sec:Applications.QKD}.

\begin{figure}[h]
    \centering
    \begin{tikzpicture}[font=\scriptsize, scale=0.95]

    \boxElement{-0.2}{4.5}{3.7}{1}{blue}{\cref{lem:reductionLemma}}{``Identity to the right"}
  
    \boxElement{-0.2}{2.5}{3.7}{1.25}{blue}{\cref{cor:variationalFormula}}{Generalized Vari- \\ational Expressions II}

    \boxElement{-0.2}{-0.25}{3.7}{1.25}{violet}{\cref{thm:PisierOriginal}}{Pisier's formula \cite{Book.Pisier.1998}}


    \boxElement{5}{3}{3.7}{1.25}{blue}{\cref{thm:general.multiplicativity}}{General orderd multiplicativity}

    \boxElement{5}{1.5}{3.7}{1.25}{blue}{\cref{thm:mainchainrule}}{$1\to(1,p)$-multiplicativity} 
    
    \boxElement{7.6}{-0.25}{3.7}{1.25}{violet}{\cref{app:Additivity.on.reduced.operator.spaces}}{Reduction argument from \cite{Himbeeck.2025}} 
    

    \boxElement{10.2}{4.5}{3.7}{1.25}{teal}{\cref{cor:Chain.Rules}}{Chain rule~\eqref{eq:chain.rule.intro}}

    \boxElement{10.2}{3}{3.7}{1.25}{teal}{\cref{cor:generalized.eat.chain.rule}}{Chain rule for tensor product maps~\eqref{eq:intro.generalized.EAT}}
      
    \boxElement{10.2}{1.25}{3.7}{1.5}{teal}{\cref{thm:IIDreduction}}{Additivity of weighted output Rényi entropy under product channels }
\draw[black, thick, ->] (1.75, 1.1) -- (1.75, 2.5);
\draw[black, thick, ->] (1.75, 3.85) -- (1.75, 4.5);
    
\draw[black, thick, ->] (3.5, 5) -- (10.2, 5);
\draw[black, thick, ->] (3.5, 4.8) -- (5, 3.5);
\draw[black, thick, ->] (3.5, 4.6) -- (5, 2.25);

\draw[black, thick, ->] (8.7, 3.5) -- (10.2, 3.5);
\draw[black, thick, ->] (8.7, 2.125) -- (10.2, 2.125);

\draw[black, thick, plus arrow]  (9.35, 1.1) -- (9.35, 2.125);

\end{tikzpicture}
    \caption{The above figure illustrates the main implications presented in this work, excluding the applications to QKD. \textcolor{violet}{Violet boxes} represent external results, \textcolor{blue}{blue boxes} our main theorems presented in terms of channel norms, and \textcolor{teal}{teal boxes} their transcription in terms of conditional Rényi-entropies.}
    \label{fig:Implications}
\end{figure}
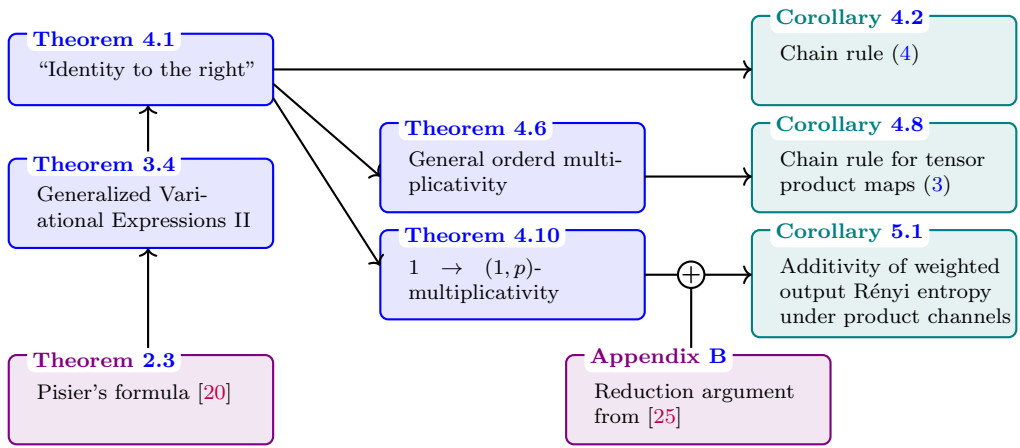

For an overview of these results and their connection, see \cref{fig:Implications}.

We note that Arqand and Tan~\cite{Arqand.2025} independently obtained similar results using different techniques.

\section{Preliminaries}\label{sec:preliminaries}

The aim of this section is to give preliminaries and set notations for this article. For \cref{sec:Generalized.Variational.Expressions} in particular we require notions of operator spaces and operator-valued Schatten norms. The required notions will be introduced in \cref{pre:operator.spaces} and \cref{pre:Operator.valued.Schatten}. There in particular we also introduce new notations for multi-index Schatten norms, which we believe to be well suited in the context of quantum information theory.

The proofs of most of the statements can be found in the main body of the text, however, proofs for either  well-known facts, or ones that are very similar to proofs in the main body are presented in the appendix.

\subsection{Basic notation}
We denote by $[n]:=\{1,...,n\}$ the set of natural numbers until $n\in\mathbb{N}$. Quantum systems are denoted by upper case latin letters $Q,R,S$, while Hilbert spaces are denoted by $\mathcal{H}$, $\mathcal{K}$, $\mathcal{H}_R$, $\mathcal{H}_1$, etc., with norm denoted e.g.~by $\|\cdot \|_{\mathcal{H}}$. They are assumed to be separable, unless explicitly stated to be finite dimensional.
Given two Hilbert spaces, we denote with $\mathcal{H}\otimes\mathcal{K}$ the Hilbert space constructed as the completion of the algebraic tensor product of these two spaces with respect to the canonical norm induced by the tensor-product inner product on $\mathcal{H}\otimes\mathcal{K}$. \\
We denote the Banach space of bounded operators from some Hilbert space $\mathcal{H}$ to some other $\mathcal{K}$, i.e. $X:\mathcal{H}\to \mathcal{K}$, as $\mathcal{B}(\mathcal{H},\mathcal{K})$, with the operator norm $\|\cdot\|_\infty$. For simplicity we write $\mathcal{B}(\mathcal{H})\equiv \mathcal{B}(\mathcal{H},\mathcal{H})$. The identity element in $\mathcal{B}(\mathcal{H})$ is denoted by $\1\equiv\1_\mathcal{H}$. More generally, we often label an operator $X$ supported on a labeled Hilbert space $\mathcal{H}_S$ as $X_S$. By slight abuse of notations, we will also denote by $X_S$ operators $X_S\otimes \1_R\in \mathcal{B}(\mathcal{H}_S\otimes\mathcal{H}_R)$ when clear from context. When $\mathcal{H}_S$ is of finite dimension, we sometimes denote its dimension by $|S|$.

An operator $X\in \mathcal{B}(\mathcal{H})$ is positive semidefinite, written $X\geq0$, if it can be written as $X=Y^*Y$ for some other operator $Y\in\mathcal{B}(\mathcal{H})$, where $Y^*$ denotes the adjoint of $Y$. The set of all positive semidefinite operators acting on some Hilbert space $\mathcal{H}$ is denoted by $\Pos(\mathcal{H})$. We denote $X>0$ if $X\geq 0$ and its kernel is trivial.

The Schatten-$p$ space over $\mathcal{H}$ with index $1\leq p\leq\infty$ is denoted by $\mathcal{S}_p(\mathcal{H})$ with associated Schatten-norm $\|X\|_p:=\Tr[|X|^p]^{\frac{1}{p}}$ when $p<\infty$ and $\|\cdot\|_\infty$ being the above mentioned operator norm, when $p=\infty$. $\mathcal{S}_p(\mathcal{K},\mathcal{H})$ is defined analogously. In both cases $\Tr[\cdot]$ is the canonical trace on $\mathcal{B}(\mathcal{H})$ and $|X|:=\sqrt{X^*X}.$ 
Note that $\mathcal{S}_\infty(\mathcal{H})$ coincides with the set of all compact operators endowed with the operator norm and that in finite dimensions we have $\mathcal{S}_\infty(\mathcal{H})=\mathcal{B}(\mathcal{H})$.

We will be denoting the partial trace as $\tr_Q[\cdot]:\mathcal{S}_1(\mathcal{H}_{QR})\to \mathcal{S}_1(\mathcal{H}_{R})$.

A trace-normalized, positive semidefinite, Schatten-$1$ operator is called a \textit{quantum state}. We will usually denote such operator with lower case greek letters $\rho,\sigma,\omega$. We denote the set of all quantum states over some Hilbert space $\mathcal{H}$ as $\mathcal{D}(\mathcal{H}):=\{\rho\in \mathcal{S}_1(\mathcal{H}
)|\Tr[\rho]=1, \rho\geq0\}$.
 
A linear map $\Phi:\mathcal{B}(\mathcal{H})\to \mathcal{B}(\mathcal{K})$ is \textit{completely positive (CP)} if $\id_{\mathcal{B}(\mathbb{C}^n)}\otimes\Phi\in\mathcal{B}(\mathbb{C}^n\otimes\mathcal{H},\mathbb{C}^n\otimes\mathcal{K})$ is a positive map for all $n\in\mathbb{N}$. It is \textit{trace preserving (TP)} if $\Tr[\Phi(X)]=\Tr[X]$ $\forall X\in \mathcal{S}_1(\mathcal{H})$.
A \textit{quantum channel} is defined as the restriction of a linear CPTP map to some state space $\mathcal{D}(\mathcal{H})$, i.e. it is an affine CPTP map $\Phi:\mathcal{D}(\mathcal{H}_Q)\to \mathcal{D}(\mathcal{H}_R)$. Its adjoint, denoted by $\Phi^*:\mathcal{B}(\mathcal{H}_R)\to \mathcal{B}(\mathcal{H}_Q)$ is a linear, CP, unital (U) map, i.e. $\Phi^*(\1_R)=\1_Q$.
We denote the identity map as $\id_S:\mathcal{B}(\mathcal{H}_S)\to \mathcal{B}(\mathcal{H}_S)$. To simplify notations, we also write $\Phi:Q\to R$ as a shorthand for the map $\Phi:\mathcal{B}(\mathcal{H}_Q)\to \mathcal{B}(\mathcal{H}_R)$.

\subsection{Operator spaces} \label{pre:operator.spaces}

In the following, we give a concise introduction into operator space theory and operator-valued Schatten norms. These will be a central tool to derive our chain rules and additivity results in \cref{sec:Generalized.Variational.Expressions} and \cref{sec:Main.Results}. 
For a more complete review of operator space theory or operator valued Schatten spaces, see \cite{Beigi.2023} or the books \cite{Book.Pisier.1998, Book.Pisier.2003}.

Operator spaces originate from the study of non-commutative geometry. They are essentially concerned with the problem of providing natural norms on spaces of vector-valued matrices and the study of the resulting structures. Since we are studying composite quantum systems, we are concerned with matrix or operator-valued matrices, which is a prime application of operator space theory. 

In the following, we let $\mathcal{X}\subset\mathcal{B}(\mathcal{K})$ be a linear subspace. Then we construct a ``natural'' family of norms on the spaces 
\begin{align}
    M_{m,n}(\mathcal{X}):=\Big\{[X_{ij}]_{i\in[m],j\in[n]}|X_{ij}\in\mathcal{X}\Big\}
\end{align} of $\mathcal{X}$-valued  $m\times n$ matrices. For simplicity we write $M_{n}(\mathcal{X})\equiv M_{n,n}(\mathcal{X})$.
We construct norms on these spaces by viewing elements of $M_{m,n}(\mathcal{X})$ as linear maps in $\mathcal{B}(\mathcal{K}^n,\mathcal{K}^m)$, where $\mathcal{K}^n:=\bigoplus_{i=1}^n\mathcal{K}=\mathbb{C}^n\otimes\mathcal{K}$, via the identification $M_{m,n}(\mathcal{X})\subset M_{m,n}(\mathcal{B}(\mathcal{K}))\simeq \mathcal{B}(\mathcal{K}^n,\mathcal{K}^m)$:
\begin{align}
    X=[X_{ij}]\in M_{m,n}(\mathcal{X}) \leftrightarrow X:\mathcal{K}^n \to \mathcal{K}^m;\quad  v^n=[v_1,\dots, v_n]^\intercal\mapsto \Big[\sum_j X_{ij}v_j\Big]^\intercal\,.
\end{align}
Hence the space $M_{m,n}(\mathcal{B}(\mathcal{K}))$ is naturally equipped with the norms induced by $\mathcal{B}(\mathcal{K}^n,\mathcal{K}^m)$, i.e.
\begin{align} 
    \|X\|_{m,n}\equiv\|X\|_{M_{m,n}(\mathcal{X})}&:=\sup\{\|Xv^n\|_{\mathcal{K}^m}|v^n\in\mathcal{K}^n, \|v^n\|_{\mathcal{K}^n}\leq 1\} \label{equ:def.mxn.norm} \\ &= \sup\left\{\left(\sum_{i=1}^m\bigg\|\sum_{j=1}^nX_{ij}v_j\bigg\|_{\mathcal{K}}^2\right)^{\frac{1}{2}}\Bigg|\sum_{j=1}^n\|v_j\|_{\mathcal{K}}^2=1\right\},
\end{align} where $\|\cdot\|_{\mathcal{K}}$ denotes the norm on $\mathcal{K}$. In the case $m=n$, we write $\|\cdot\|_n:=\|\cdot\|_{n,n}$. 
For simplicity we will also denote $\|X\|_{\mathcal{X}}\equiv \|X\|_{M_1(\mathcal{X})}$.
\begin{proposition} \label{prop:central.properties.OSnorms}
The family of norms defined above satisfies the following two main properties, namely for any $m,n\in\mathbb{N}$,
\begin{align}
\text{(i)} \quad & \|FXG\|_m \leq \|F\|_{m,n}\|X\|_n\|G\|_{n,m} \quad \forall F,G^*\in M_{m,n}(\mathbb{C}), X\in M_n(\mathcal{X}), \\
\text{(ii)} \quad & \|X\oplus Y\|_{m+n} = \max\{\|X\|_n,\|Y\|_m\} \quad \forall X\in M_n(\mathcal{X}), Y\in M_m(\mathcal{X}),
\end{align} where 
$X\oplus Y= \begin{pmatrix}
    X & 0 \\
    0 & Y
\end{pmatrix}$, and $FXG\equiv (F\otimes \1_\mathcal{K})X(G\otimes \1_\mathcal{K})$.
\end{proposition}
More generally,
\begin{definition} A linear space $\mathcal{X}$ with a family of norms $\|\cdot \|_{m,n}$ on $M_{m,n}(\mathcal
X)$ that satisfy the above properties (i) and (ii) is called an \textit{(abstract) operator space}.
\end{definition}

It turns out that the only linear spaces $\mathcal{X}$ with endowed norms $\|\cdot \|_{m,n}$ on $M_{m,n}(\mathcal{X})$ satisfying properties (i) and (ii) are closed linear subspaces of some $\mathcal{B}(\mathcal{K})$, where $\mathcal{K}$ is a (possibly infinite dimensional) Hilbert space. 
Hence closed linear subspaces $\mathcal{X}\subset\mathcal{B}(\mathcal{K})$ are called \textit{(concrete) operator spaces.}

\subsection{Norms on operator-valued Schatten spaces}\label{pre:Operator.valued.Schatten}

Next we describe the operator space structure of \textit{operator-valued Schatten norms}, sometimes also referred to as \textit{amalgamated} $L_p$ \textit{norms} or \textit{Pisier norms}. For simplicity of introduction we let $\mathcal{H}$ be a Hilbert space of dimension $d<\infty$ here. Given some operator space $\mathcal{X}$, we set
\begin{align}
\mathcal{S}_\infty[\mathcal{H},\mathcal{X}]:=M_d(\mathcal{X}),
\end{align} i.e. the $d\times d$ matrices taking values in $\mathcal{X}$, with the norm $\|\cdot\|_d\equiv\|\cdot\|_{M_{d}(\mathcal{X})}$. It is an operator space since we have 
\begin{align}
M_{m,n}(\mathcal{S}_\infty[\mathcal{H},\mathcal{X}]) \simeq M_{md,nd}(\mathcal{X}),
\end{align} which clearly satisfy \cref{prop:central.properties.OSnorms} $(i),(ii)$.
We remark here that $\mathcal{S}_\infty[\mathcal{H},\mathbb{C}]=\mathcal{S}_\infty(\mathcal{H})$ follows directly from the definition above.

The main technical tool in this work are the operator space norms of Schatten-$q$-spaces of $\mathcal{X}$-valued operators $\mathcal{S}_p[\mathcal{H},\mathcal{X}]$. Since a complete description of the operator space structure of $\mathcal{S}_p[\mathcal{H},\mathcal{X}]$ is beyond the scope of the present paper, we refer to \cite{Beigi.2023,Devetak.2006} for more details. For their original construction via interpolation between certain Haagerup-tensor products of row- and column-operator spaces, see \cite{Book.Pisier.1998}.
Remarkably, we can omit this because we are able to define and work with these norms by only understanding the norm on $\mathcal{S}_\infty[\mathcal{H},\mathcal{X}]$ discussed above. Next, we enumerate some of their core properties.
 
In the case where $\mathcal{H}$ is finite dimensional, these operator-valued Schatten spaces should be thought of as linear spaces of $d\times d$ matrices valued in $\mathcal{X}$, with special norms $\|\cdot\|_{\mathcal{S}_q[\mathcal{H},\mathcal{X}]}$. These extend naturally to  infinite dimensional settings.
Importantly, in the case where $\mathcal{X}=\mathbb{C}$ they all reduce to the well-known Banach space of Schatten class operators \cite{Beigi.2023}, i.e.
\begin{align}
   \mathcal{S}_q[\mathcal{H},\mathbb{C}]= \mathcal{S}_q(\mathcal{H}).
\end{align} 
They also satisfy the following duality relation, namely
\begin{align}
\label{eq:duality-op-space}
   \left( \mathcal{S}_p[\mathcal{H},\mathcal{X}]\right)^* = \mathcal{S}_{p^\prime}[\mathcal{H},\mathcal{X}^*],
\end{align} where $\frac{1}{p}+\frac{1}{p^\prime}=1$ are dual indices and $\mathcal{X}^*$ is the operator space dual of $\mathcal{X}$ \cite{Book.Pisier.2003}.
Since the norms for other values of $q$ are defined via interpolation between $\mathcal{S}_\infty[\mathcal{H},\mathcal{X}]$ and $\mathcal{S}_1[\mathcal{H},\mathcal{X}]$ (which can be defined via~\eqref{eq:duality-op-space}) \cite{Book.Pisier.1998}, they satisfy many desirable properties, including Pisier's formula, which we take as their definition. 

\begin{theorem}[Pisier's formula \cite{Book.Pisier.1998}] \label{thm:PisierOriginal}
Let $\mathcal{H}$ be a separable Hilbert space and $\mathcal{X}$ an operator space. Then for $1\leq p\leq\infty$ the following variational formulas hold for any $X\in \mathcal{S}_p[\mathcal{H},\mathcal{X}]$.
\begin{align}
    \|X\|_{\mathcal{S}_p[\mathcal{H},\mathcal{X}]}= \inf_{\underset{X=FYG}{F,G\in \mathcal{S}_{2p}(\mathcal{H}), Y\in \mathcal{S}_\infty[\mathcal{H},\mathcal{X}]}}\|F\|_{2p}\|Y\|_{\mathcal{S}_\infty[\mathcal{H},\mathcal{X}]}\|G\|_{2p}.
\end{align}
\end{theorem}
Firstly we note that it is not mandatory that $F,G$ are square, i.e. the statement still holds when taking the infimum over $F,G,Y$ s.t. $F,G^*\in\mathcal{S}_{2p}(\mathcal{K},\mathcal{H})$ and $Y\in\mathcal{S}_\infty[\mathcal{K},\mathcal{X}]$ as long as $\dim\mathcal{H}\leq \dim\mathcal{K}$. The statement in the case when $\mathcal{H}=\mathcal{K}$ is however sufficient in practice. Next, we prove two important direct consequences of Pisier's formula.
\begin{corollary}\label{cor:Pisier.Formula}
1) Let $\mathcal{X}$ be an operator space. Then, the norm on $S_p[\mathcal{H},\mathcal{X}]$ is invariant under local isometries $V^*,U\in\mathcal{B}(\mathcal{H},\mathcal{K})$ satisfying $U^*U=VV^*=\1_\mathcal{H}$. That is for any $X\in\mathcal{S}_{p}[\mathcal{H},\mathcal{X}]$ and any such $V^*,U\in\mathcal{B}(\mathcal{H},\mathcal{K})$ it holds that 
\begin{align}
    \|UXV\|_{S_p[\mathcal{K},\mathcal{X}]} =    \|X\|_{S_p[\mathcal{H},\mathcal{X}]}.
\end{align}
In particular, when $\mathcal{K}=\mathcal{H}$, this means invariance under unitaries on the first system. \\
2) When $\dim\mathcal{H}<\infty$ one can restrict the infimum effectively over positive semidefinite operators, and hence
\begin{align}
\|X\|_{\mathcal{S}_p[\mathcal{H},\mathcal{X}]}=\inf_{F,G\in\mathcal{S}_{2p}(\mathcal{H}), F,G\geq0}\|F\|_{2p}\|G\|_{2p}\|F^{-1}XG^{-1}\|_{\mathcal{S}_\infty[\mathcal{H},\mathcal{X}]}=
\inf_{\underset{\|F\|_1=\|G\|_1=1}{F,G\ge 0}}\|F^{-\frac{1}{2p}}XG^{-\frac{1}{2p}}\|_{\mathcal{S}_\infty[\mathcal{H},\mathcal{X}]}\,,
\end{align}
where $F^{-1},G^{-1}$ denote the generalized (Moore-Penrose) inverses of $F,G$. 
\end{corollary}

\begin{proof}
The first claim follows via isometric invariance of the Schatten norms,
\begin{align}
\|UXV\|_{\mathcal{S}_p[\mathcal{K},\mathcal{X}]} &= \inf_{\underset{UXV=FYG}{F,G\in \mathcal{S}_{2p}(\mathcal{K}), Y\in \mathcal{S}_\infty[\mathcal{K},\mathcal{X}]}}\|F\|_{2p}\|Y\|_{\mathcal{S}_\infty[\mathcal{K},\mathcal{X}]}\|G\|_{2p} \\ &= \inf_{\underset{X=(U^*F)Y(GV^*)}{F,G\in \mathcal{S}_{2p}(\mathcal{K}), Y\in \mathcal{S}_\infty[\mathcal{K},\mathcal{X}]}}\|F\|_{2p}\|Y\|_{\mathcal{S}_\infty[\mathcal{K},\mathcal{X}]}\|G\|_{2p}  \\ &= \inf_{\underset{X=F^\prime YG^\prime}{F^\prime,G^{\prime*}\in \mathcal{S}_{2p}(\mathcal{K},\mathcal{H}), Y\in \mathcal{S}_\infty[\mathcal{K},\mathcal{X}]}}\|F^\prime\|_{2p}\|Y\|_{\mathcal{S}_\infty[\mathcal{K},\mathcal{X}]}\|G^\prime\|_{2p} \\ 
 &= \|X\|_{\mathcal{S}_p[\mathcal{H},\mathcal{X}]},
\end{align} where in the third line we redefined $F^\prime=U^*F\in\mathcal{S}_{2p}(\mathcal{K},\mathcal{H})$ and used that they have identical Schatten-$2p$-norms. The fourth line follows from the comment above on non-square $F,G$.

\medskip

\noindent To prove the second statement, we let $X\in\mathcal{S}_{p}[\mathcal{H},\mathcal{X}]$, assuming $\dim\mathcal{H}<\infty$. Then, similarly to above, observe that
\begin{align}
\|X\|_{\mathcal{S}_p[\mathcal{H},\mathcal{X}]} &= \inf_{\underset{F,G,Y  \text{ s.t. } X=FYG}{F,G\in \mathcal{S}_{2p}(\mathcal{H}), Y\in \mathcal{S}_\infty[\mathcal{H},\mathcal{X}]}}\|F\|_{2p}\|Y\|_{\mathcal{S}_\infty[\mathcal{H},\mathcal{X}]}\|G\|_{2p} \\     
&= \inf_{\underset{F,G,Y  \text{ s.t. }X=P_F(UYV)P_G}{F,G\in \mathcal{S}_{2p}(\mathcal{H}), Y\in \mathcal{S}_\infty[\mathcal{H},\mathcal{X}]}}\|P_F\|_{2p}\|Y\|_{\mathcal{S}_\infty[\mathcal{H},\mathcal{X}]}\|P_G\|_{2p} \\ 
&= \inf_{\underset{F,G,Y  \text{ s.t. }X=P_FYP_G}{F,G\in \mathcal{S}_{2p}(\mathcal{H}), Y\in \mathcal{S}_\infty[\mathcal{H},\mathcal{X}]}}\|P_F\|_{2p}\|Y\|_{\mathcal{S}_\infty[\mathcal{H},\mathcal{X}]}\|P_G\|_{2p},
\end{align} where in the second line we set $F=P_FU$ and $G=VP_G$ to be the right-, respectively, left-polar decompositions of $F,G$, such that $P_F,P_G\geq 0$. For the third equality we used the above and renamed $UYV$ to $Y$. Denote with $P_F^{-1},P_G^{-1}$ their Moore-Penrose inverses and with $\Pi_F:=P_FP_F^{-1}=P_F^{-1}P_F$ the projection onto the support of $P_F$, and analogously for $G$. 
Now for a triple $(F,G,Y)$ that occurs in the infimum we define $\tilde{Y}:=\Pi_FY\Pi_G$. Then by \cref{prop:central.properties.OSnorms} $i)$, see also \cite[Lemma 1.6]{Book.Pisier.1998}, it follows that
\begin{align}
\|\Pi_FY\Pi_G\|_{\mathcal{S}_\infty[\mathcal{H},\mathcal{X}]} \leq \|Y\|_{\mathcal{S}_\infty[\mathcal{H},\mathcal{X}]},
\end{align} since $\|\Pi_F\|,\|\Pi_G\|\leq 1$.
Hence it holds that
\begin{align}
    \|X\|_{\mathcal{S}_p[\mathcal{H},\mathcal{X}]} \geq
    \inf_{\underset{F,G,Y  \text{ s.t. }X=P_FYP_G}{F,G\in \mathcal{S}_{2p}(\mathcal{H}), Y\in \mathcal{S}_\infty[\mathcal{H},\mathcal{X}]}}\|P_F\|_{2p}\|\Pi_FY\Pi_G\|_{\mathcal{S}_\infty[\mathcal{H},\mathcal{X}]}\|P_G\|_{2p}.
\end{align}
On the other hand for any suitable triple $(F,G,Y)$ it follows that
\begin{align}
    X=P_FYP_G=P_F\Pi_FY\Pi_GP_G=P_F\tilde{Y}P_G,
\end{align} hence $(F,G,\Pi_FY\Pi_G)$ is also a compatible triple. Since we have by definition an injection from suitable triples $(F,G,Y)$ to ones $(F,G,\Pi_FY\Pi_G)$ it follows that
\begin{align}
     \|X\|_{\mathcal{S}_p[\mathcal{H},\mathcal{X}]} \leq
    \inf_{\underset{  X=P_FYP_G}{F,G\in \mathcal{S}_{2p}(\mathcal{H}), Y\in \mathcal{S}_\infty[\mathcal{H},\mathcal{X}]}}\|P_F\|_{2p}\|\Pi_FY\Pi_G\|_{\mathcal{S}_\infty[\mathcal{H},\mathcal{X}]}\|P_G\|_{2p}.
\end{align}
So overall we have shown equality. Now by construction we further have
\begin{align}
  P_F^{-1}XP_G^{-1} = \Pi_FY\Pi_G=\tilde{Y} 
\end{align} and in total we get
\begin{align}
   \|X\|_{\mathcal{S}_p[\mathcal{H},\mathcal{X}]} &= \inf_{P_F,P_G\in \mathcal{S}_{2p}(\mathcal{H}), P_F,P_G\geq 0}\|P_F\|_{2p}\|P_F^{-1}XP_G^{-1}\|_{\mathcal{S}_\infty[\mathcal{H},\mathcal{X}]}\|P_G\|_{2p} \\
&=\inf_{\underset{\|F\|_1=\|G\|_1=1}{F,G\ge 0}}\|F^{-\frac{1}{2p}}XG^{-\frac{1}{2p}}\|_{\mathcal{S}_\infty[\mathcal{H},\mathcal{X}]}\,,
\end{align} which is what we wanted to show.
\end{proof}

Although it is not obvious from the above, the expression in \cref{thm:PisierOriginal} does define a norm. In particular, it satisfies the triangle inequality, H\"{o}lder's duality, and the property that $\|X\|_{\mathcal{S}_q[\mathcal{H},\mathcal{S}_q(\mathcal{K})]}=\|X\|_{\mathcal{S}_q(\mathcal{H}\otimes \mathcal{K})}$.
This last property follows from the more general fact that for two Hilbert spaces $\mathcal{H}, \mathcal{K}$
\begin{align} \label{equ:combining.spaces}
\mathcal{S}_q[\mathcal{H},\mathcal{S}_q[\mathcal{K},\mathcal{X}]] \simeq \mathcal{S}_q[\mathcal{H}\otimes\mathcal{K},\mathcal{X}]\simeq \mathcal{S}_q[\mathcal{K},\mathcal{S}_q[\mathcal{H},\mathcal{X}]], \end{align} where $\simeq$ means they are completely isomorphic, see \cref{subsec:Norms.on.linear.spaces}, i.e. equal as operator spaces \cite[Theorem 1.9]{Book.Pisier.1998}.

Pisier's \cref{thm:PisierOriginal} was also used to give tractable variational expressions for the norms on the spaces $\mathcal{S}_q[\mathcal{H},\mathcal{S}_p(\mathcal{K})]$, see e.g. the case $\mathcal{X}=\mathcal{S}_p[\mathcal{K},\mathbb{C}]=\mathcal{S}_p(\mathcal{K})$ \cite[Section 3.5]{Devetak.2006}.
These make the operator-valued Schatten norms with two indices very tractable for applications in quantum information theory, see e.g. \cite{Beigi.2016,Bardet.2022,Beigi.2023, Cheng.2024, Devetak.2006, Gao.2023, Gupta.2015, Wilde.2014}. 

\begin{theorem}[\cite{Book.Pisier.1998}]\label{thm:2.pisier.variational.expressions}
\label{thm:expression-pq}
Given an element $X\in \mathcal{S}_q[\mathcal{H}_1,\mathcal{S}_p(\mathcal{H}_2)]$ acting on the Hilbert space $\mathcal{H}_1\otimes\mathcal{H}_2$ it holds that
\begin{align}
\|X_{12}\|_{\mathcal{S}_q[\mathcal{H}_1,\mathcal{S}_p(\mathcal{H}_2)]}
=
    \begin{cases}
        \inf_{\underset{X_{12}=F_1Y_{12}G_1}{F,G\in \mathcal{S}_{2r}(\mathcal{H}_1),Y\in \mathcal{S}_{p}(\mathcal{H}_{12})}} \|F\|_{2r}\|G\|_{2r}\|Y\|_{p}\,, & \quad \textup{for } q\leq p \\
        \|X_{12}\|_p\,, & \quad \textup{for } q=p \\
        \sup_{F,G\in\mathcal{S}_{2r}(\mathcal{H}_1)} \|F\|^{-1}_{2r}\|G\|^{-1}_{2r}\|F_1X_{12}G_1\|_p, & \quad \textup{for } q\geq p
    \end{cases}
\end{align}
where the infimum and supremum are over $F,G\in \mathcal{S}_{2r}(\mathcal{H}_1)$ acting only on the first Hilbert space and $Y\in \mathcal{S}_q(\mathcal{H}_{12})$ with $\frac{1}{r}:=\left|\frac{1}{q}-\frac{1}{p}\right|$.
Further if $X_{12}\geq0$, then one can choose $F=G$, see \cite{Devetak.2006} or \cite[Proposition 3.1 (v,vi)]{Bardet.2022} and \cite[Proposition 5.2 iii)]{Bardet.2024}.
\end{theorem}
Some of their central properties are summarized in \cite{Devetak.2006}.
In \cref{thm:VariationalSpOperatorspace} below, we show that one can also obtain a version of these for the more general case where $\mathcal{S}_q(\mathcal{H})$ is replaced by any other operator space $\mathcal{X}$.

\begin{notation}[multi-index Schatten norms]
We introduce the following notation to keep track of the Schatten indices occurring in the norms, their order, value, and associated quantum systems.
Given an operator $X\in \mathcal{S}_q[\mathcal{H}_A,\mathcal{S}_p[\mathcal{H}_B,\mathcal{S}_r(\mathcal{H}_C)]]$ acting on the tripartite quantum system $\mathcal{H}_A\otimes\mathcal{H}_B\otimes\mathcal{H}_C$ we write 
\begin{align}
    \|X\|_{(A:q,B:p,C:r)} \equiv \|X\|_{\mathcal{S}_q[\mathcal{H}_A,\mathcal{S}_p[\mathcal{H}_B,\mathcal{S}_r(\mathcal{H}_C)]]}
\end{align} 
and more generally for a suitable $X\in\mathcal{B}(\otimes_{i=1}^k\mathcal{H}_{A_i})$
\begin{align}
    \|X\|_{(A_1:q_1...\,A_k:q_k)} \equiv \|X\|_{\mathcal{S}_{q_1}[\mathcal{H}_{A_1}...\,\mathcal{S}_{q_k}(\mathcal{H}_{A_k})...]}.
\end{align} 
Likewise, for an operator $X\in \mathcal{S}_q[\mathcal{H}_A,\mathcal{S}_p[\mathcal{H}_B,\mathcal{X}]]$ we will write
\begin{align}
    \|X\|_{(A:q,B:p;\mathcal{X})} \equiv \|X\|_{\mathcal{S}_q[\mathcal{H}_A,\,\mathcal{S}_p[\mathcal{H}_B,\mathcal{X}]]}
\end{align} and analogously for a different number of indices. 
Importantly, note also that the order in which the systems appear is determined by the order of the indices in the norm, and whenever possible we will try to make the order of systems in the operator match.

That is, formally we have in this notation for $X\in\mathcal{B}(\mathcal{H}_A)$ and $Y\in\mathcal{B}(\mathcal{H}_B)$,
\begin{align}
    \|X\otimes Y\|_{(A:q,B:p)} =  \|Y\otimes X\|_{(A:q,B:p)} = \|X\|_q\|Y\|_p\,, 
 \end{align}
 but we try to avoid the notation $\|Y\otimes X\|_{(A:q,B:p)}$ as much as possible to avoid confusions.
\end{notation}
The following is a direct consequence of \eqref{equ:combining.spaces}:
\begin{proposition}
\label{prop:combining.systems}
 Consecutive Schatten indices of the same value can be combined: 
\begin{align}
\|X\|_{(A_1:q_1,...,A_{m-1}:q_{m-1},A_{m}:p,...,A_{m+n}:p;\mathcal{X})} = \|X\|_{(A_1:q_1...,A_{m-1}:q_{m-1},A_{m}\cdots  A_{m+n}:p;\mathcal{X})},
\end{align} 
This in particular implies that if all Schatten indices are equal, then the norm
\begin{align}
    \|X\|_{(A_1:p...\,A_k:p)}=\|X\|_p
\end{align} reduces to the Schatten-$p$-norm of $X$.
\end{proposition}

In the next proposition, we show that operator-valued Schatten norms enjoy a natural multiplicativity property.

\begin{proposition}
\label{prop:spliting.systems.left.}
For any $A\otimes X\in S_q[\mathcal{H},\mathcal{X}]$,
\begin{align}
    \|A\otimes X\|_{S_q[\mathcal{H},\mathcal{X}]}=\|A\|_q\|X\|_{\mathcal{X}}\,.
\end{align}
 More generally for $\bigotimes_{i=1}^k X_i\in \mathcal{S}_{q_1}[\mathcal{H}_{A_1}[\mathcal{S}_{q_2}[...\,\mathcal{S}_{q_k}(\mathcal{H}_{A_k})]...]]$,
\begin{align} 
\bigg\|\bigotimes_{i=1}^k X_i\bigg\|_{(A_1:q_1,...,A_k:q_k)}= \prod_{i=1}^k \|X_i\|_{(A_i:q_i)}.
\end{align}
\end{proposition}
\begin{proof}
In case that $d:=\dim\mathcal{H}<\infty$ the first statement follows from \cref{cor:Pisier.Formula}: 
\begin{align}
    \|A\otimes X\|_{S_q[\mathcal{H},\mathcal{X}]} &= \inf_{F,G\in\mathcal{S}_{2q}(\mathcal{H}), F,G\geq0}\|F\|_{2q}\|G\|_{2q}\|F^{-1}AG^{-1}\otimes X\|_{\mathcal{S}_\infty[\mathcal{H},\mathcal{X}]} \\
    &=\inf_{F,G\in\mathcal{S}_{2q}(\mathcal{H}), F,G\geq0}\|F\|_{2q}\|G\|_{2q}\|F^{-1}AG^{-1}\|_\infty\|X\|_{\mathcal{X}} \\
    &= \|A\|_q\|X\|_\mathcal{X}.
\end{align} 
Here in the second line we used that $\|B\otimes X\|_{S_\infty[\mathcal{H},\mathcal{X}]}=\|B\otimes X\|_{M_d(\mathcal{X})}=\|B\|_{M_d(\mathbb{C})}\|X\|_\mathcal{X}= \|B\|_\infty\|X\|_\mathcal{X}$, which follows from \cref{prop:central.properties.OSnorms} and the definition of the $M_d(\mathcal{X})$ norm. 
Alternatively this also follows by assuming without loss of generality that $A$ is block-diagonal, since else we may absorb SVD-unitaries into the norm by \cref{cor:Pisier.Formula}, and applying \cite[Corollary 1.3]{Book.Pisier.1998}.
The second statement now follows directly by induction:
\begin{align}
   \bigg\|\bigotimes_{i=1}^k X_i\bigg\|_{(A_1:q_1,...,A_k:q_k)} &=   \bigg\|X_1\otimes \bigotimes_{i=2}^{k} X_i\bigg\|_{S_{q_1}[\mathcal{H}_{A_1},\,\mathcal{X}]} 
   = \|X_1\|_{q_1} \bigg\|\bigotimes_{i=2}^k X_i\bigg\|_{(A_2:q_2,...,A_k:q_k)},
\end{align} where we used $\mathcal{X}=S_{q_2}[\mathcal{H}_{A_2}...\,\mathcal{S}_{q_k}(\mathcal{H}_{A_k})]...]$.
\end{proof}

\subsection{Norms on linear maps}\label{subsec:Norms.on.linear.spaces}
For a linear map $\Phi: \mathcal{S}_{q_1}[\mathcal{H}_{A_1}...\,\mathcal{S}_{q_k}(\mathcal{H}_{A_k})...]\to \mathcal{S}_{p_1}[\mathcal{H}_{B_1}\,...\,\mathcal{S}_{p_l}(\mathcal{H}_{B_l})...]$ we set 
\begin{align}
    \|\Phi\|_{(A_1:q_1...\,A_k:q_k)\to (B_1:p_1...\,B_l:p_l)}:=\sup_{X\neq 0} \frac{\|\Phi(X)\|_{(B_1:p_1...\,B_l:p_l)}}{\|X\|_{(A_1:q_1...\,A_k:q_k)}},
\end{align} where the supremum is over all $X\in \mathcal{S}_{q_1}[\mathcal{H}_{A_1}...\,\mathcal{S}_{q_k}(\mathcal{H}_{A_k})...]$.
We write a $^+$ superscript next to its channel norm when we restrict the optimization over positive semidefinite operators, i.e. 
\begin{align}
     \|\Phi\|^+_{(A_1:q_1...\,A_k:q_k)\to (B_1:p_1...\,B_l:p_l)}:=\sup_{X\geq 0, X\neq0} \frac{\|\Phi(X)\|_{(B_1:p_1...\,B_l:p_l)}}{\|X\|_{(A_1:q_1...\,A_k:q_k)}}.
\end{align} 
This is useful when taking these maps as quantum channels acting on (positive) quantum states. For most channel norms of interest in our applications it is known that for a CP map $\Phi$, one can restrict the optimization over positive elements only without changing the norm, i.e. $\|\Phi\|=\|\Phi\|^+$, see 
\cref{lem:positivesufficiency}.

The \textit{completely bounded (cb)} norm of a linear map is defined as
\begin{align}
    \|\Phi\|_{cb,(A_1:q_1...\,A_k:q_k)\to (B_1:p_1...\,B_l:p_l)} := \sup_E\|\id_E\otimes\Phi\|_{(E:\infty,A_1:q_1...\,A_k:q_k)\to (E:\infty,B_1:p_1...\,B_l:p_l)}\,,
\end{align}
where the supremum is over environment systems $E$ of arbitrary size. It was shown in \cite[Lemma 1.7]{Book.Pisier.1998}, however, that the cb-norm is independent of the index of the environment $E$, 
\begin{align}
     \|\Phi\|_{cb,(A_1:q_1...\,A_k:q_k)\to (B_1:p_1...\,B_l:p_l)} = \sup_E\|\id_E\otimes\Phi\|_{(E:t,A_1:q_1...\,A_k:q_k)\to (E:t,B_1:p_1...\,B_l:p_l)},
\end{align} for any $1\leq t\leq \infty$. We will be choosing this index at our convenience.
One immediate consequence of this is that for any linear map $\Phi:\mathcal{X}\to \mathcal{Y}$ between operator spaces it holds that
\begin{align}
\|\id_A\otimes\Phi\|_{cb,\mathcal{S}_q[\mathcal{H}_A,\mathcal{X}]\to \mathcal{S}_q[\mathcal{H}_A,\mathcal{Y}]} = \|\Phi\|_{cb,\mathcal{X}\to \mathcal{Y}} 
\end{align} for any quantum system $A$. \\
A linear map between operator spaces is called a \textit{complete isometry} if it is invertible and the map and  its inverse both have a CB norm equal to $1$.
A well-known occurrence of the CB norm in quantum information theory is the diamond norm: for a difference of channels $\Phi-\Psi:A\to B$
\begin{align}
\|\Phi-\Psi\|_\diamond = \sup_E\,\|\id_E\otimes (\Phi-\Psi)\|_{(EA:1)\to (EB:1)} 
= \|\Phi-\Psi\|_{cb,(A:1)\to (B:1)}\,. 
\end{align}

\medskip

\noindent 
Of important interest to the present discussion is the SWAP map $F_{A\leftrightarrow B}:\mathcal{B}(\mathcal{H}_A\otimes\mathcal{H}_B)\to \mathcal{B}(\mathcal{H}_B\otimes\mathcal{H}_A): X_A\otimes X_B\mapsto X_B\otimes X_A$. It is is a $\textit{complete contraction}$ when acting on $\mathcal{S}_p[\mathcal{H}_A,\mathcal{S}_q(\mathcal{H}_B)]$ for $q\geq p$ \cite[Theorem 8]{Devetak.2006}. This means that for $q\geq p$, any $t$ and system $E$:
\begin{align} \label{equ:complete.contraction}
      \|\id_E\otimes F_{A \leftrightarrow B}\|_{(E:t,A:p,B:q)\to (E:t,B:q,A:p)}
    \leq \|F_{A \leftrightarrow B}\|_{cb,(A:p,B:q)\to (B:q,A:p)} \leq 1\,.
\end{align}
This implies that, in our notations $\|X\|_{(B:q,A:p)}=\|X_{BA}\|_{(B:q,A:p)}\leq \|X_{AB}\|_{(A:p,B:q)}=\|X\|_{(A:p,B:q)}$, whenever $q\geq p$. Further due to \eqref{equ:combining.spaces} it holds that for any operator space $\mathcal{X}$ and quantum systems $E,A,B,$
\begin{align}\label{equ:SWAP.isometry}
    \|\id_E\otimes F_{A\leftrightarrow B}\otimes\id_{\mathcal{X}}\|_{(E:t,A:q,B:q;\mathcal{X})\to (E:t,B:q,A:q;\mathcal{X})} = 1
\end{align} is a complete isometry for any $q$.


The CB norm as defined above satisfies nice properties, which makes it very versatile and powerful. One of them is a very simple general chain rule, which we make use of multiple times.

\begin{lemma}\label{lem:cb.norm.splitting}
Let $\Psi:\mathcal{X}\to\mathcal{Y}$ and $\Phi:\mathcal{Y}\to\mathcal{Z}$ be two maps between arbitary operator spaces $\mathcal{X},\mathcal{Y},\mathcal{Z}$, then
\begin{align}
\|\Phi\circ\Psi\|_{cb,\mathcal{X}\to\mathcal{Z}} \leq \|\Psi\|_{cb,\mathcal{X}\to \mathcal{Y}}\cdot \|\Phi\|_{cb,\mathcal{Y}\to \mathcal{Z}}
\end{align}
\end{lemma}


\subsection{R\'{e}nyi conditional entropies}\label{renyidefs}

We emphasize an important consequence of the variational expression that makes it a powerful tool in quantum information theory. Namely that by the first variational expression in \cref{thm:2.pisier.variational.expressions} one may express the  optimized R\'{e}nyi-conditional entropy as a $\|\cdot\|_{(1,\alpha)}$ norm.  
Recall that for $1< \alpha<\infty$ the sandwiched R\'{e}nyi$-\alpha$ divergence \cite{Lennert.2013, Wilde.2014} between two quantum states $\rho,\sigma\in\mathcal{D}(\mathcal{H})$ is defined as
\begin{align}
    D_\alpha(\rho\|\sigma):= \frac{\alpha}{\alpha-1}\log\|\sigma^{\frac{1-\alpha}{2\alpha}}\rho\sigma^{\frac{1-\alpha}{2\alpha}}\|_\alpha
\end{align}
and the optimized R\'{e}nyi$-\alpha$ conditional entropy is defined as
\begin{align}
    H^\uparrow_\alpha(X|Y)_\rho := -\inf_{\sigma_Y\in \mathcal{D}(\mathcal{H}_Y)}D_{\alpha}(\rho_{XY}\|\1_X\otimes\sigma_Y). 
\end{align} 
In the limit $\alpha\to 1^+$, we recover the standard conditional entropy $H_1(A|B)_\rho\equiv H(A|B)_\rho$. In \cite[(3.22)]{Devetak.2006}, without yet calling it as such, the above was expressed as the following $(1,\alpha)$-Pisier norm,
\begin{align}
   H^\uparrow_\alpha(X|Y)_\rho = \frac{\alpha}{1-\alpha}\log\|\rho_{YX}\|_{(Y:1,X:\alpha)}.
\end{align} 

Similarly we have that, given a quantum channel (a CPTP map) $\Phi:Q\to RS$, the minimum $\alpha$-R\'{e}nyi entropy of the output system $S$ given $R$ under this map is proportional to the logarithm of the channel $1\to(1,p)$ norm, since
\begin{align}
    \inf_{\rho_{Q}\in\mathcal{D}(\mathcal{H}_{Q})}H_\alpha^\uparrow(S|R)_{\Phi(\rho)} &= \inf_{\rho\in\mathcal{D}(\mathcal{H}_{Q})}\frac{\alpha}{1-\alpha}\log\|\Phi(\rho)\|_{(R:1,S:\alpha)} \\ &= \frac{\alpha}{1-\alpha}\log \sup_{\rho \geq 0} \frac{\|\Phi(\rho)\|_{(R:1,S:\alpha)}}{\| \rho \|_{(Q:1)}}  \\ &= \frac{\alpha}{1-\alpha}\log\|\Phi\|^+_{(Q:1)\to(R:1,S:\alpha)}.
    \label{eq:operator_norm_min_entropy}
\end{align}

For the $cb,1\to(1,p)$ norm we have a similar connection. Namely, its logarithm is related to the minimal conditional R\'{e}nyi entropy under purifications, i.e.
\begin{align} \label{eq:cb.1.to.1p.norm}
\inf_{E}\inf_{\rho_{EQ}}H^\uparrow_\alpha(S|RE)_{\id_E\otimes\Phi(\rho)} = \inf_{\rho_{Q}}H^\uparrow_\alpha(S|R\tilde{Q})_{(\id_{\tilde{Q}}\otimes\Phi)(|\sqrt{\rho}\rangle\langle\sqrt{\rho}|_{\tilde{Q}Q})}  = \frac{\alpha}{1-\alpha}\log\|\Phi\|^+_{cb,(Q:1)\to (R:1,S:\alpha)},
\end{align}
where the infimum in both cases is over positive normalized states $\rho$ and $\tilde{Q}$ are systems isomorphic to $Q$. Here $|\sqrt{\rho}\rangle\langle\sqrt{\rho}|_{\tilde{Q}Q}$ is a purification of $\rho_Q$ with purifying system $\tilde{Q}$, e.g. $|\sqrt{\rho}\rangle:=\sum_{i}|i\rangle_{\tilde{Q}}\otimes\sqrt{\rho_Q}|i\rangle_{Q}$.
This follows from a standard Schmidt decomposition argument that yields that the CB norm is achieved on environments $E$ which are isomorphic to the input system $Q$ of the channel, see \cref{lem:finiteEnvironement} for details.

In the next section, we derive useful variational expressions for the norms on the operator spaces $\mathcal{S}_q[\mathcal{H}_Q,\mathcal{S}_p[\mathcal{H}_P,\mathcal{S}_r(\mathcal{H}_R)]]$, i.e. for operator-valued Schatten norms of 3 indexes, departing from \cref{thm:PisierOriginal}. 
These will be central to show chain rules and sub-multiplicativity of completely bounded $1\to (1,p)$ norms that will translate directly into additivity statements for conditional  entropies under tensor products of quantum channels.

\section{Generalized variational expressions for Pisier norms}\label{sec:Generalized.Variational.Expressions}

In this section, we first derive generalized 
variational formulas for relating the norms of $\mathcal{S}_p[\mathcal{H},\mathcal{X}]$ and $\mathcal{S}_q[\mathcal{H},\mathcal{X}]$ in \cref{thm:VariationalSpOperatorspace}. By iteration, this will allow us to derive tractable formulas for the norms of multi-index Schatten norms $\mathcal{S}_{p_1}[\cH_{A_1},\mathcal{S}_{p_2}[...\,\mathcal{S}_{p_k}(\cH_{A_k})\dots]]$. In particular we derive variational formulas for systems made of three subsystems in \cref{cor:variationalFormula}. 

Pisier's formula \cref{thm:PisierOriginal} gives a way of relating the $p$-norm on the ``left'' most system to the $\infty$-norm on it. In the following, we refine it to relate the $p$-norm to its $q$-norm, as done in \cref{thm:2.pisier.variational.expressions}. 
Here, we keep $\mathcal{X}$ general, so the following result generalizes \cref{thm:2.pisier.variational.expressions}, which can be recovered in the special case $\mathcal{X}=\mathcal{S}_q(\mathcal{K})$. Although the proof closely follows that of the mentioned special case, we provide it here for completeness.


\begin{lemma}[Generalized variational expressions]
Let $\mathcal{X}$ be an operator space. Then for $1\leq p\leq q\leq \infty$ and $\frac{1}{r}=\frac{1}{p}-\frac{1}{q}$, the following variational formulas hold for any $X\in \mathcal{S}_p[\mathcal{H},\mathcal{X}],\mathcal{S}_q[\mathcal{H},\mathcal{X}]$ respectively:
\begin{align}
\operatorname{(i)}\qquad&\|X\|_{\mathcal{S}_p[\mathcal{H},\mathcal{X}]}=\inf_{\substack{F,G\in \mathcal{S}_{2r}(\mathcal{H}), Y\in\mathcal{S}_q[\mathcal{H}, \mathcal{X}] \\X=FYG}} \|F\|_{2r} \|Y\|_{\mathcal{S}_q[\cH,\mathcal{X}]}\|G\|_{2r}\\
\operatorname{(ii)}\qquad &\|X\|_{\mathcal{S}_q[\mathcal{H},\mathcal{X}]} = \sup_{F,G\in \mathcal{S}_{2r}(\mathcal{H})}\|F\|_{2r}^{-1}\|FXG\|_{\mathcal{S}_p[\mathcal{H},\mathcal{X}]}\|G\|_{2r}^{-1}\,.
\end{align} \label{thm:VariationalSpOperatorspace}
\end{lemma}
\noindent The power of this theorem is that it can be iterated by choosing $\cX$ itself to be composed of many subsystems with Schatten spaces with potentially different indices.

\begin{proof}
We will prove (i) by separately showing upper and lower bounds and (ii) will then follow by duality. 
Note that to keep notations short and as done before, we omit writing identity operators, e.g. for $F,G\in \mathcal{S}_{2r}(\mathcal{H})$ and $Y\in \mathcal{S}_\infty[\mathcal{H},\mathcal{X}]$ we will write $FYG\equiv(F\otimes\1)Y(G\otimes\1)$. 

    (i) Let $X=FYG\in \mathcal{S}_p[\mathcal{H},\mathcal{X}]$ with $F,G\in \mathcal{S}_{2r}(\mathcal{H}),Y\in \mathcal{S}_q[\mathcal{H},X]$ and suppose $Y=HZK$ with $H,K\in \mathcal{S}_{2q}(\mathcal{H}), Z\in \mathcal{S}_\infty[\mathcal{H},{\mathcal{X}}]$, then $X=FH Z KG$ and by H\"{o}lder's inequality $\|FH\|_{2p}\leq \|F\|_{2r}\|H\|_{2q}$, since $\frac{1}{p}=\frac{1}{r}+\frac{1}{q}$, we get $FH,KG\in \mathcal{S}_{2p}(\mathcal{H})$. Hence
    \begin{align}
\|X\|_{\mathcal{S}_p[\mathcal{H},\mathcal{X}]}&\leq \|FH\|_{2p}\|Z\|_{\mathcal{S}_\infty[\mathcal{H},\mathcal{X}]}\|KG\|_{2p} \\&\leq \|F\|_{2r}\|H\|_{2q}\|Z\|_{\mathcal{S}_\infty[\mathcal{H},\mathcal{X}]}\|K\|_{2q}\|G\|_{2r}\,.
    \end{align} 
     After minimization over $H,K,Z$, we obtain
    \begin{align}
\|X\|_{\mathcal{S}_p[\mathcal{H},\mathcal{X}]} \leq \|F\|_{2r}\|Y\|_{\mathcal{S}_q[\mathcal{H},\mathcal{X}]}\|G\|_{2r} \implies  \|X\|_{\mathcal{S}_p[\mathcal{H},\mathcal{X}]} \leq \inf_{\substack{F,G\in \mathcal{S}_{2r}\\Y\in \mathcal{S}_q[\mathcal{H},\mathcal{X}]}}\|F\|_{2r}\|Y\|_{\mathcal{S}_q[\mathcal{H},\mathcal{X}]}\|G\|_{2r}\,.
    \end{align}

    On the other hand, let $\epsilon>0$. Then there exists $Y\in \mathcal{S}_\infty[\mathcal{H},\mathcal{X}]$ and $F,G\in \mathcal{S}_{2p}(\mathcal{H})$ such that $X=FYG$ and 
    \begin{align}
\|X\|_{\mathcal{S}_p[\mathcal{H},\mathcal{X}]}+\epsilon \geq \|F\|_{2p}\|Y\|_{\mathcal{S}_\infty[\mathcal{H},\mathcal{X}]}\|G\|_{2p}\,.
    \end{align}
    By performing a polar decomposition of $F$ and $G$ and absorbing the unitaries into $Y$, which does not change the norm, we may assume $F,G$ to be positive semidefinite and thus $F^{\frac{p}{q}},G^{\frac{p}{q}}\in \mathcal{S}_{2q}(\mathcal{H})$ and $F^{\frac{p}{r}},G^{\frac{p}{r}}\in \mathcal{S}_{2r}(\mathcal{H})$. As a result, $X = F^{\frac{p}{r}} F^{\frac{p}{q}} Y G^{\frac{p}{q}} G^{\frac{p}{r}}$. Hence

    \begin{align}
        \inf_{\substack{X=HWK\\H,K\in \mathcal{S}_{2r}(\mathcal{H}), W\in \mathcal{S}_q[\mathcal{H},\mathcal{X}], }}\|H\|_{2r}\|W\|_{\mathcal{S}_q[\mathcal{H},\mathcal{X}]}\|K\|_{2r} &\leq \|F^{\frac{p}{r}}\|_{2r}\|F^{\frac{p}{q}}YG^{\frac{p}{q}}\|_{\mathcal{S}_q[\mathcal{H},\mathcal{X}]}\|G^{\frac{p}{r}}\|_{2r} \\
        &\leq \|F^{\frac{p}{r}}\|_{2r}\|F^{\frac{p}{q}}\|_{2q}\|Y\|_{\mathcal{S}_\infty[\mathcal{H},\mathcal{X}]}\|G^{\frac{p}{q}}\|_{2q}\|G^{\frac{p}{r}}\|_{2r} \\&=\|F\|_{2p}\|Y\|_{\mathcal{S}_\infty[\mathcal{H},\mathcal{X}]} \|G\|_{2p} = \|X\|_{\mathcal{S}_p[\mathcal{H},\mathcal{X}]}+\epsilon.
    \end{align}
    Since $\epsilon>0$ is arbitrary the claim follows.

    \medskip 

   (ii) by the duality $\mathcal{S}_{q}[\mathcal{H},\mathcal{X}]^*=\mathcal{S}_{q^\prime}[\mathcal{H},\mathcal{X}^*]$, i.e. there is a complete isometry between these two operator spaces, where $1/q+1/q^\prime=1$ (see \cite[Proposition 4.3]{Beigi.2023}). For $1/p + 1/p^\prime = 1$, we have $1 \leq p^\prime \leq q^\prime$ and also $\frac{1}{r} = \frac{1}{p^\prime} - \frac{1}{q^\prime}$, we thus obtain
    \begin{align}
    \|X\|_{\mathcal{S}_q[\mathcal{H},\mathcal{X}]}
        &=\sup_Y\frac{|\Tr[Y^*X]|}{\|Y\|_{\mathcal{S}_{q\prime}[\mathcal{H},\mathcal{X}^*]}}\\
        &\overset{\operatorname{(i)}}{=} \sup_Y\sup_{ Y = FZG}\frac{|\Tr[Y^*X]|}{\|F\|_{2r}\|G\|_{2r}\|Z\|_{\mathcal{S}_{p^\prime}[\mathcal{H},\mathcal{X}^*]}} \\ &=\sup_{F,G,Z}\frac{|\Tr[G^*Z^*F^*X]|}{\|F\|_{2r}\|G\|_{2r}\|Z\|_{\mathcal{S}_{p^\prime}[\mathcal{H},\mathcal{X}^*]}}\\
        &=\sup_{F,G,Z}\frac{|\Tr[Z^*(F^*XG^*)]|}{\|F\|_{2r}\|G\|_{2r}\|Z\|_{\mathcal{S}_{p^\prime}[\mathcal{H},\mathcal{X}^*]}}\\
&=\sup_{F,G}\frac{\|FXG\|_{\mathcal{S}_p[\mathcal{H},\mathcal{X}]}}{\|F\|_{2r}\|G\|_{2r}}.
    \end{align}
\end{proof}
Since for our applications we are mainly interested in operator-valued Schatten norms over $3$ indices, we derive explicit variational formulas for the latter from the above lemma. However, in principle the same approach leads one to explicit variational formulas for arbitrarily many different indices. This is the main technical result of this section.

\begin{theorem}[Variational formulas 1] \label{cor:variationalFormula.2}
    Consider $X_{123}\in \mathcal{S}_p[\mathcal{H}_1,\mathcal{S}_q[\mathcal{H}_2,\mathcal{S}_s(\mathcal{H}_3)]]$. For $1\leq p\leq q\leq s\leq \infty$ with $\frac{1}{r}:=\frac{1}{p}-\frac{1}{s}$ and $ \frac{1}{r^\prime}:=\frac{1}{q}-\frac{1}{s}$, it holds that
  \begin{align}
\|X\|_{(1:p,2:q,3:s)} = \inf_{ X=G_{12}Y_{123}F_{12}}\|GG^*\|^{\frac{1}{2}}_{(1:r,2:r^\prime)}\|F^*F\|^{\frac{1}{2}}_{(1:r,2:r^\prime)}\|Y\|_s\,.
  \end{align}
For $1\leq s\leq q\leq p\leq \infty$ with $\frac{1}{r}:=\frac{1}{s}-\frac{1}{p}$ and $\frac{1}{r^\prime}:=\frac{1}{s}-\frac{1}{q}$, 
  \begin{align}
\|X\|_{(1:p,2:q,3:s)} = \sup_{G_{12},F_{12}}\|G^*G\|^{-\frac{1}{2}}_{(1:r,2:r^\prime)}\|FF^*\|^{-\frac{1}{2}}_{(1:r,2:r^\prime)}\|G_{12}X_{123}F_{12}\|_s.
  \end{align}  
\end{theorem}

These two expressions should be thought of as generalizations of \cref{thm:2.pisier.variational.expressions} to three subsystems, for the above specified order of indices.
A direct consequence of the Theorem is that the multi-index Piser norm of tensor product operators splits on the right, as compared to \cref{prop:spliting.systems.left.}:
\begin{corollary}
Let $X\in\mathcal{S}_{p}[\mathcal{H}_1,\mathcal{S}_q[\mathcal{H}_2,\mathcal{S}_s(\mathcal{H}_3)]]$ s.t. $X=Y_{12}\otimes Z_3$, then if either $1\leq p\leq q\leq s\leq \infty,$ or $1\leq s\leq q\leq p\leq \infty,$ it holds that
\begin{align}
    \|Y_{12}\otimes Z_3\|_{(1:p,2:q,3:s)} = \|Y\|_{(1:p,2:q)}\cdot\|Z\|_s
\end{align}
\end{corollary}
\begin{proof}
The proof follows from the multiplicativity of the Schatten-$s$-norm and the fact that relating the $(p,q,s)$ to the $(s,s,s)$ norm can be done by only affecting the first two systems. 
Depending on the order of $p,q,s$ we apply the corresponding variational formula from \cref{cor:variationalFormula.2}. In the case of $1\leq s\leq q\leq p\leq \infty$ we get
\begin{align}
   \|Y_{12}\otimes Z_3\|_{(1:p,2:q,3:s)} &= \sup_{G_{12},F_{12}}\|G^*G\|^{-\frac{1}{2}}_{(1:r,2:r^\prime)}\|FF^*\|^{-\frac{1}{2}}_{(1:r,2:r^\prime)}\|G_{12}Y_{12}F_{12}\otimes Z_3\|_s \\&= \sup_{G_{12},F_{12}}\|G^*G\|^{-\frac{1}{2}}_{(1:r,2:r^\prime)}\|FF^*\|^{-\frac{1}{2}}_{(1:r,2:r^\prime)}\|G_{12}Y_{12}F_{12}\|_s \cdot\|Z_3\|_s \\ &= \|Y\|_{1:p,2:q}\|Z\|_s.
\end{align}
In the case $1\leq p\leq q\leq s$ we can by an argument as in \cref{cor:Pisier.Formula} assume $G,F\geq 0$ and apply the same as above now to $\|G_{12}^{-1}Y_{12}F^{-1}_{12}\otimes Z_3\|_s=\|G_{12}^{-1}Y_{12}F^{-1}_{12}\|_{s}\cdot\|Z_3\|_s$.
\end{proof}

In a similar fashion to \cref{cor:variationalFormula.2} we can also generalize Pisier's formula and \cite[Equation (3.3)]{Devetak.2006}. We do this in the following.

\begin{theorem}[Variational Formulas 2]\label{cor:variationalFormula}
Consider $X_{123}\in \mathcal{S}_p[\mathcal{H}_1,\mathcal{S}_q[\mathcal{H}_2,\mathcal{X}]]$ for any operator space $\mathcal{X}$, then for any $1\leq p,q\leq \infty$ it holds that
\begin{align}
       \|X\|_{(1:p,2:q;\mathcal{X})}\leq\inf_{ X=G_{12}Z_{123}F_{12}}\|GG^*\|^{\frac{1}{2}}_{(1:p,2:q)}\|F^*F\|^{\frac{1}{2}}_{(1:p,2:q)}\|Z\|_{(1:\infty,2:\infty;\mathcal{X})}, 
\end{align} where $(1:q,2:p;\mathcal{X})$ is a shorthand for the norm on $\mathcal{S}_q[\mathcal{H}_1,\mathcal{S}_p[\mathcal{H}_2,\mathcal{X}]]$.
Further for $1\leq p\leq q\leq \infty$ equality holds.
\end{theorem}

\begin{proof}[Proof of \cref{cor:variationalFormula.2}]
The proof consists of repeated applications of \cref{thm:VariationalSpOperatorspace} combined with simplifications due to \cref{thm:2.pisier.variational.expressions}. The first formula follows from \cref{thm:VariationalSpOperatorspace} (i), first with $\mathcal{X} = \mathcal{S}_q[\mathcal{H}_2,\mathcal{S}_s(\mathcal{H}_3)]$ and then with $\mathcal{X} = \mathcal{S}_s(\mathcal{H}_3)$:

\begin{align}
\notag \|X\|_{(1:p,2:q,3:s)} &= \inf_{X_{123} =F_1Y_{123}G_1} \|F\|_{2\alpha}\|G\|_{2\alpha}\|Y\|_{(1:q,2:q,3:s)} \\
&= \inf_{X_{123} =F_1Y_{123}G_1} \|F\|_{2\alpha}\|G\|_{2\alpha}\|Y\|_{(12:q,3:s)} 
\\ &= \inf_{ X_{123}=F_1 H_{12} Z_{123} K_{12} G_1} \|F\|_{2\alpha} \|G\|_{2\alpha} \|H\|_{2r^\prime} \|K\|_{2r^\prime} \|Z\|_s 
\\&= \inf_{\substack{X_{123}=F_1H_{12}Z_{123}K_{12}G_1\\F_1, G_1\geq 0,H_{12},K_{12} }} \|F\|_{2\alpha} \|G\|_{2\alpha} \|HH^*\|^{\frac{1}{2}}_{r^\prime} \|K^*K\|^{\frac{1}{2}}_{r^\prime} \|Z\|_s 
\\ &= \inf_{\substack{X_{123}=M_{12} Z_{123} N_{12}\\F_1,G_1\geq 0 }} \|F\|_{2\alpha} \|G\|_{2\alpha} \|F_1^{-1}M_{12}M_{12}^*F_1^{-1}\|^{\frac{1}{2}}_{r^\prime} \|G_1^{-1}N_{12}^*N_{12}G_1^{-1}\|^{\frac{1}{2}}_{r^\prime} \|Z\|_s
\\ &= \inf_{ X_{123}=M_{12}Z_{123}N_{12}}\|M_{12}M_{12}^*\|^{\frac{1}{2}}_{(1:r,2:r^\prime)}\|N_{12}^*N_{12}\|^{\frac{1}{2}}_{(1:r,2:r^\prime)}\|Z\|_s
    \end{align}
    where in the fourth line we restricted to positive $F,G$ by polar decomposition, absorbed unitaries into $H,K$ respectively, and set $M\equiv FH$, $N\equiv KG$ and thus $H=F^{-1}M$, $K=NG^{-1}$, where these inverses respectively are the generalized Moore-Penrose inverses of $F,G$. There we also used that the Schatten norms of $H^*H$ and $HH^*$ are equal, since their non-zero singular values, which are equal to their eigenvalues, are equal. 
    In the last line, we used the expression of $\|\cdot\|_{(1:r,2:r^\prime)}$ given in~\cref{thm:expression-pq} and the fact that $\frac{1}{\alpha}=\frac{1}{r}-\frac{1}{r^\prime}=\frac{1}{p}-\frac{1}{q}$. We recall that since $M_{12}M_{12}^*$ and $N^*_{12}N_{12}$ are positive, their 2-indexed Schatten norms are can be written by optimizations over just one matrix-variable $F,G$, respectively.

\medskip 

The second formula follows analogously after applying Lemma \ref{thm:VariationalSpOperatorspace} (ii). We 
define $\frac{1}{\alpha}=\frac{1}{q}-\frac{1}{p}$.
\begin{align}
&\|X\|_{(1:p,2:q,3:s)} \\
&\qquad = \sup_{F_1,G_1} \|F\|^{-1}_{2\alpha}\|G\|^{-1}_{2\alpha}\|F_1 X_{123}G_1\|_{(1:q,2:q,3:s)} \\  
&\qquad =\sup_{\substack{0\leq F_1,G_1\\H_{12},K_{12}}}\|F\|^{-1}_{2\alpha}\|G\|^{-1}_{2\alpha}\|H\|^{-1}_{2r^\prime}\|K\|^{-1}_{2r^\prime}  \|H_{12}F_1X_{123}G_1K_{12}\|_s \\
&\qquad = \sup_{\substack{0\leq F_1,G_1\\ M_{12},N_{12} }}\|F\|^{-1}_{2\alpha}\|G\|^{-1}_{2\alpha}\|F_1^{-1}M_{12}^*M_{12}F_1^{-1}\|^{-\frac{1}{2}}_{2r^\prime}\|G_1^{-1}N_{12}N^*_{12}G_1^{-1}\|^{-\frac{1}{2}}_{r^\prime}  \|M_{12}X_{123}N_{12}\|_s \\ 
&\qquad =\!\!\!\sup_{ M_{12},N_{12}}\!\!\!\left(\inf_{0<F_1}\!\!\!\|F\|^2_{2\alpha}\|F_1^{-1}M_{12}^*M_{12}F_1^{-1}\|_{r^\prime}\!\!\right)^{\!\!\!-\frac{1}{2}}\!\!\left(\inf_{0<G_1}\!\!\!\|G\|^2_{2\alpha}\|G_1^{-1}N_{12}N_{12}^*G^{-1}_1\|_{r^\prime}\!\!\right)^{\!\!\!\!-\frac{1}{2}} \, \!\!\|M_{12}X_{123}N_{12}\|_s \\ 
&\qquad = \sup_{M_{12},N_{12}}\|M^*M\|^{-\frac{1}{2}}_{(1:r,2:r^\prime)}\|FF^*\|^{-\frac{1}{2}}_{(1:r,2:r^\prime)} \|M_{12}X_{123}N_{12}\|_s.
\end{align} 
\end{proof}

Similarly we prove the other variational expression.
\begin{proof}[Proof of \cref{cor:variationalFormula}]
We split the proof into the cases $p\leq q$ and $p\geq q$.
In the first case the claimed formula follows analogously to the first one in \cref{cor:variationalFormula.2}, when using Theorem \ref{thm:PisierOriginal} instead of Lemma \ref{thm:VariationalSpOperatorspace} $(i)$.
Given $\frac{1}{\alpha}=\frac{1}{p}-\frac{1}{q}$, we get
\begin{align}
\|X\|_{(1:p,2:q;\mathcal{X})} &{=} \inf_{ X=F_1Y_{123}G_1} \|F\|_{2\alpha}\|G\|_{2\alpha}\|Y\|_{(1:q,2:q;\mathcal{X})} \\ &= \inf_{X=F_1H_{12}Z_{123}K_{12}G_1} \|F\|_{2\alpha} \|G\|_{2\alpha} \|H\|_{2q} \|K\|_{2q} \|Z\|_{(1:\infty,2:\infty;\mathcal{X})} \\ &=  \inf_{ X=M_{12}Z_{123}N_{12}}\|MM^*\|^{\frac{1}{2}}_{(1:p,2:q)}\|N^*N\|^{\frac{1}{2}}_{(1:p,2:q)}\|Z\|_{(1:\infty,2:\infty;\mathcal{X})},
\end{align} 
where we used Lemma \ref{thm:VariationalSpOperatorspace} (i) in the first line above, and skipped some steps since they are identical to those in the proof of the first formula above upon replacing $r^\prime\leftrightarrow q$ and keeping the last operator space $\mathcal{X}$ instead of $\mathcal{S}_s(\mathcal{H}_3)$. The inequality in the setting $p\geq q$ follows similarly as the above. Setting $\frac{1}{r}=\frac{1}{q}-\frac{1}{p}$, we have
\begin{align}
&\|X\|_{(1:p,2:q;\mathcal{X})} \\
&\qquad = \sup_{F_1,G_1}\|F\|_{2r}^{-1}\|G\|_{2r}^{-1}\|F_1X_{123}G_1\|_{(1:q,2:q;\mathcal{X})} \\
&\qquad= \sup_{F_1,G_1}\|F\|_{2r}^{-1}\|G\|_{2r}^{-1} \inf_{F_1X_{123}G_1=H_{12}Z_{123}K_{12}}\|H\|_{2q}\|K\|_{2q}\|Z\|_{(1:\infty,2:\infty;\mathcal{X})} \\
&\qquad = \sup_{F_1,G_1\ge 0}\|F\|_{2r}^{-1}\|G\|_{2r}^{-1} \inf_{ F_1X_{123}G_1=H_{12}Z_{123}K_{12}}\|H\|_{2q}\|K\|_{2q}\|Z\|_{(1:\infty,2:\infty;\mathcal{X})} \\
&\qquad = \sup_{F_1,G_1\ge  0}\|F\|_{2r}^{-1}\|G\|_{2r}^{-1} \inf_{ X_{123}=F^{-1}_1H_{12}Z_{123}K_{12}G_1^{-1}}\|HH^*\|^{\frac{1}{2}}_{q}\|K^*K\|^{\frac{1}{2}}_{q}\|Z\|_{(1:\infty,2:\infty;\mathcal{X})}
    \\
    &\qquad = \sup_{F_1,G_1\ge 0}\|F\|_{2r}^{-1}\|G\|_{2r}^{-1} \inf_{ X_{123}=M_{12}Z_{123}N_{12}}\!\!\!\!\|F_1M_{12}M^*_{12}F_1\|^{\frac{1}{2}}_{q}\|G_1N^*_{12}N_{12}G_1\|^{\frac{1}{2}}_{q}\|Z\|_{(1:\infty,2:\infty;\mathcal{X})}
    \\ &\qquad \leq \inf_{X_{123}=M_{12}Z_{123}N_{12}}\sup_{F_1\ge 0}\|F\|_{2r}^{-1}\|F_1M_{12}M^*_{12}F_1\|^{\frac{1}{2}}_{q}\!\sup_{G_1\ge  0}\|G\|_{2r}^{-1}\|G_1N^*_{12}N_{12}G_1\|^{\frac{1}{2}}_{q}\!\|Z\|_{(1:\infty,2:\infty;\mathcal{X})}
    \\&\qquad = \inf_{ X_{123}=M_{12}Z_{123}N_{12}}\|MM^*\|^{\frac{1}{2}}_{(1:p,2:q)}\|N^*N\|^{\frac{1}{2}}_{(1:p,2:q)}\|Z\|_{(1:\infty,2:\infty;\mathcal{X})},
\end{align} 
where the inequality arises from switching the supremum and infimum. 
\end{proof}


Having established these variational formulas, we now turn our attention to applying them to prove our main  results about the multiplicativity of certain (completely bounded) mixed operator norms under tensor products, including in particular $1\to (1,p)$ norms.

\section{Main results: Chain rules and additivity}\label{sec:Main.Results}

We now use the operator-valued Schatten- and Pisier-norms introduced in the previous sections to establish useful properties of entropic measures evaluated on a composite system. We start with a chain rule that allows decomposing the entropy of a joint system $ST$ into the appropriate entropies of $S$ and $T$. 

\subsection{Chain Rules}

Our chain rule will follow from the following multiplicativity statement stating that an identity ``on the right'' does not affect the norm of CP maps.
\begin{theorem} \label{lem:reductionLemma}
  Let $\Phi$ be a CP map $\Phi:QP\to RS$, $\mathcal{X}$ an operator space and $1\leq q\leq p\leq \infty,1\leq r,s\leq \infty$ then
    \begin{align}        \|\Phi\otimes\id_{\mathcal{X}}\|_{(Q:q,P:p;\mathcal{X})\to(R:r,S:s;\mathcal{X})} = \|\Phi\|^+_{(Q:q,P:p)\to (R:r,S:s)},
    \end{align} where the superscript $^+$ denotes optimization over positive semidefinite operators. 
\end{theorem}

This result is a generalization of \cite[Lemma 5]{Devetak.2006}. In fact,~\cite[Lemma 5]{Devetak.2006} corresponds to the special case where $P$ and $S$ are trivial and our proof strategy closely follows the one in~\cite{Devetak.2006}.  

Before giving the proof of this result, we discuss some consequences.
A first immediate one is that
\begin{align}
    \|\Phi\|_{(Q:q,P:p)\to(R:r,S:s)} =     \|\Phi\|^+_{(Q:q,P:p)\to(R:r,S:s)}, 
\end{align}
for the above specified indices.
A second direct consequence are two chain rules expressed in terms of conditional R\'enyi entropies. The first one is
\begin{corollary}[Amortized Chain rule]\label{cor:amortized.chain.rule}
For any state $\rho\in\mathcal{D}(\rho_{QPT})$ on system $QPT$ and any quantum channel $\Phi:QP\to RS$ we have the following chain rule for $\alpha > 1$
\begin{align}
    H_\alpha^\uparrow(ST|R)_{(\Phi\otimes\id_T)(\rho)} - H_\alpha^\uparrow(PT|Q)_{\rho} \geq \inf_{\omega\in\cD(\cH_{QP})} \left( H_\alpha^\uparrow(S|R)_{\Phi(\omega)}-H^\uparrow_\alpha(P|Q)_\omega \right)
\end{align}
\end{corollary}

The second one is
\begin{corollary}[Chain rule]\label{cor:Chain.Rules}
For any state $\rho \in \mathcal{D}(\cH_{QT})$ on systems $QT$ and any quantum channel $\Phi:Q\to RS$, we have the chain rule for $\alpha > 1$
\begin{align}
\label{eq:chain-rule-Q}
H^{\uparrow}_\alpha(ST|R)_{(\Phi 
\otimes \id_T)(\rho_{QT})} - H^{\uparrow}_\alpha(T|Q)_\rho &\geq   \inf_{\sigma \in \mathcal{D}(\cH_{Q})} H^{\uparrow}_\alpha(S|R)_{\Phi(\sigma_{Q})},
\end{align} where the inequality is saturated for density operators $\rho_{QT}=\rho_{Q}\otimes \rho_T$ where $\rho_Q$ achieves the infimum on the right-hand side expression. 
\end{corollary}

\begin{remark}
This second chain rule~\eqref{eq:chain-rule-Q} should be compared to the one in~\cite[Lemma 3.6]{Metger.2024} with the replacements $Q \to E$, $\emptyset \to R$, $S \to A'$, $T \to A$, $R \to E'$. Our chain rule is less general in the sense that the system $R$ in~\cite[Lemma 3.6]{Metger.2024} is chosen to be trivial (this is the relevant setting for the analysis of prepare-and-measure protocols~\cite{Metger.2023}), but it has some advantages/differences: firstly, it is directly for the optimized R\'enyi entropy, secondly, we obtain a slight improvement in that we do not need a purifying system on the right-hand side of~\eqref{eq:chain-rule-Q},  and thirdly, we have no loss in the parameter $\alpha$. 
\end{remark}

\begin{proof}[Proof of \cref{cor:amortized.chain.rule} and \cref{cor:Chain.Rules}]
\noindent They follow both directly from \cref{lem:reductionLemma} by setting $\mathcal{X}=S_\alpha(\mathcal{H}_T)$ and applying $\frac{\alpha}{1-\alpha}\log$. The amortized one follows when setting $q=r=1, p=s=\alpha> 1$, whereas \eqref{eq:chain-rule-Q} by setting $P$ trivial and  $p=q=r=1, s=\alpha> 1$. Then the left-hand sides become, respectively, with or without the $P$
\begin{equation}
    \inf_{\rho_{QPT}\geq 0}\left( H^{\uparrow}_\alpha(ST|R)_{(\Phi \otimes \id_T)(\rho_{QPT})} - H^{\uparrow}_\alpha(PT|Q)_{\rho_{QT}}\right)
\end{equation} and right-hand side, 
\begin{equation}
    \inf_{\sigma_{QP}\geq 0} \left(H_\alpha^\uparrow(S|R)_{\Phi(\sigma)}-H^\uparrow_\alpha(P|Q)_\sigma\right),
\end{equation} respectively, 
\begin{equation}
    \inf_{\sigma \in \mathcal{D}(\mathcal{H}_Q)}  H^{\uparrow}_\alpha(S|R)_{\Phi(\sigma_{Q})}.
\end{equation} 
Due to positive homogeneity the suprema over $\sigma\geq 0$ and $\sigma\in\cD(\cH_Q)$ are equal.
\end{proof}

We now proceed with the proof of \cref{lem:reductionLemma}, which as previously mentioned closely follows \cite[Lemma 5]{Devetak.2006} up to using our \cref{cor:variationalFormula} in place of \cref{thm:PisierOriginal}.
\begin{proof}[Proof of \cref{lem:reductionLemma}]
First of all notice that $\|\Phi\otimes\id_{\mathcal{X}}\|_{(Q:q,P:p;\mathcal{X})\to(R:r,S:s;\mathcal{X})} \geq \|\Phi\|^+_{(Q:q,P:p)\to (R:r,S:s)}$, since we can just restrict the supremum on the left hand side over product operators. In fact, consider $X_{QP}\otimes Y_{\mathcal{X}}$, then by \cref{thm:VariationalSpOperatorspace} and multiplicativity of the operator-valued Schatten norms \cref{prop:spliting.systems.left.}, we obtain both
\begin{align} 
\|(\Phi\otimes\id_{\mathcal{X}})(X_{QP}\otimes Y_{\mathcal{X}})\|_{(R:r,S:s;\mathcal{X})} &= \|\Phi(X_{QP})\otimes Y_{\mathcal{X}}\|_{(R:r,S:s;\mathcal{X})}= \|\Phi(X_{QP})\|_{(R:r,S:s)}\cdot \|Y_{\mathcal{X}}\|_{\mathcal{X}}, \\ 
\|X_{QP}\otimes Y_\mathcal{X}\|_{(Q:q,P:p;\mathcal{X})} &= \|X\|_{(Q:q,P:p)}\cdot\|Y_\mathcal{X}\|_{\mathcal{X}}.
\end{align}

Now let $\1_\mathcal{X}$ be the identity element of $\mathcal{X}$.
To prove the non-trivial side of the inequality, inspired by \cite[Proof of Lemma 5]{Devetak.2006}, we consider a Kraus representation of $\Phi$: $\Phi(\rho)=\sum^\nu_{i=1}K_i \rho K_i^*$. Then for a given $\rho\in\mathcal{S}_q[\mathcal{H}_Q,\mathcal{S}_p[\mathcal{H}_P,\mathcal{X}]]$, there exist for any $\epsilon>0$, by our extension of Pisier's formula in \cref{cor:variationalFormula} for this norm with $q\leq p$, operators $A,B,Y$, s.t. $\rho=(A_{QP}\otimes\1_\mathcal{X})Y(B_{QP}\otimes \1_\mathcal{X})$ and $ \|\rho\|_{(Q:q,P:p;\mathcal{X})}\geq \|AA^*\|^{\frac{1}{2}}_{(Q:q,P:p)}\|B^*B\|^{\frac{1}{2}}_{(Q:q,P:p)}\|Y\|_{(Q:\infty,P:\infty;\mathcal{X})}-\epsilon$.
We have \begin{align}
    (\Phi\otimes\id_\mathcal{X})(\rho) = \sum_{i=1}^\nu(K_i\otimes\1_\mathcal{X})\rho(K_i^*\otimes\1_\mathcal{X})=\sum_{i=1}^\nu(K_iA\otimes\1_\mathcal{X})Y(BK_i^*\otimes\1_\mathcal{X}) = V_A(\1_{\mathbb{C}^\nu}\otimes Y)V^*_B,
\end{align} where $V_A= (K_1A\otimes\1_\mathcal{X},K_2A\otimes\1_\mathcal{X},...,K_\nu A\otimes\1_\mathcal{X})$ is a block row-vector with blocks $K_iA\otimes \1_\mathcal{X}$, and $V_B^*$ a block-column vector with blocks $BK_i^*\otimes\1_\mathcal{X}$, where we denote with $\nu$ the number of blocks and with $N$ the system in which these block live and on which $\1_{\mathbb{C}^\nu}$ acts.  
These operators $V_A, V_B$ can be embedded into the space $\mathcal{B}(\mathbb{C}^\nu\otimes\mathcal{H}_Q\otimes\mathcal{H}_P\otimes\mathcal{X},\mathbb{C}^\nu\otimes\mathcal{H}_R\otimes\mathcal{H}_S\otimes\mathcal{X})$ by padding suitably with rows of $0$ operators. For $V^*_A, V^*_B$ similarly into $\mathcal{B}(\mathbb{C}^\nu\otimes\mathcal{H}_R\otimes\mathcal{H}_S\otimes\mathcal{X},\mathbb{C}^\nu\otimes\mathcal{H}_Q\otimes\mathcal{H}_P\otimes\mathcal{X})$, by padding with columns of $0$ operators. Call these extended operators, respectively, $V_A^\prime, V_B^\prime, V_A^{\prime*} , V_B^{\prime*}$. We get $V^\prime_A=\sum^{\nu}_{i,j=1}\delta_{i1}|i\rangle\langle j|_N\otimes K_jA\otimes\1_\mathcal{X} \in \mathcal{B}(\mathbb{C}^\nu\otimes\mathcal{H}_Q\otimes\mathcal{H}_P\otimes\mathcal{X},\mathbb{C}^\nu\otimes\mathcal{H}_R\otimes\mathcal{H}_S\otimes\mathcal{X})$. 
Hence it holds that $V^\prime_AV_A^{\prime *}=|1\rangle\langle1|_N\otimes V_AV_A^*$, similarly for $B$ and $V^\prime_A(\1_N\otimes Y)V_B^{\prime *}=|1\rangle\langle 1|_N\otimes (\Phi\otimes\id_{\cX})(\rho)$. 
Now using \cref{cor:variationalFormula} on the space $\mathcal{S}_r[\mathbb{C}^\nu\otimes\mathcal{H}_R,\mathcal{S}_s[\mathcal{H}_S,\mathcal{X}]]$ we get
\begin{align}
    \|(\Phi\otimes\id_{\cX})(\rho_{QP\cX})\|_{(R:r,S:s;\mathcal{X})} &= \||1\rangle\langle 1|_N\otimes(\Phi\otimes\id_\mathcal{X})(\rho_{QPT})\|_{(N:r,R:r,S:s;\mathcal{X})} \\ 
    &= \|V^\prime_A(\1_N\otimes Y)V_B^{\prime *}\|_{(NR:r,S:s;\mathcal{X})}  
    \\ &\leq \|V^\prime_AV_A^{\prime *}\|^{\frac{1}{2}}_{(NR:r,S:s)}\|V^\prime_BV_B^{\prime *}\|^{\frac{1}{2}}_{(NR:r,S:s)}\|\1_N\otimes Y\|_{(N:\infty,Q:\infty,P:\infty;\mathcal{X})} \\ &= \|\sum_iK_iAA^*K_i^*\|^{\frac{1}{2}}_{(R:r,S:s)}\|\sum_iK_iB^*BK_i^*\|^{\frac{1}{2}}_{(R:r,S:s)}\|Y\|_{(Q:\infty,P:\infty;\mathcal{X})} \\&= \|\Phi(AA^*)\|^{\frac{1}{2}}_{(R:r,S:s)}\|\Phi(B^*B)\|^{\frac{1}{2}}_{(R:r,S:s)} \|Y\|_{(Q:\infty,P:\infty;\mathcal{X})} \\&\leq  \|\Phi\|^+_{(Q:q,P:p)\to(R:r,S:s)}\|AA^*\|^{\frac{1}{2}}_{(Q:q,P:p)}\|B^*B\|^{\frac{1}{2}}_{(Q:q,P:p)}\|Y\|_{(Q:\infty,P:\infty;\mathcal{X})}
    \\&\leq \|\Phi\|^+_{(Q:q,P:p)\to(R:r,S:s)}(\|\rho\|_{(Q:q,P:p;\mathcal{X})}+\epsilon).
\end{align} In the first and third equality, we used the multiplicativity of the operator-valued Schatten norms~\cref{prop:spliting.systems.left.}.
In the second line, we combined systems of equal indices. Since $\epsilon$ was arbitrary the claim follows. 
\end{proof} 
In the following we will extend this to completely bounded norms in the following way.

\begin{corollary} \label{Cor:rightIdentity}
    Let $\Phi:Q \to RS$ be a CP map, then for any $1\leq q,p\leq\infty$
    \begin{align}
        \|\Phi\otimes\id_{\cX}\|_{cb,(Q:q,\cX)\to (R:q,S:p,\cX)}= \|\Phi\|^+_{cb,Q:q\to (R:q,S:p)}.
    \end{align}
\end{corollary}
\begin{proof}
We have, using the above \cref{lem:reductionLemma} that 
    \begin{align}
        \|\Phi\otimes\id_{\cX}\|_{cb,(Q:q;\cX)\to (R:q,S:p;\cX)} &= \sup_E\|\underbrace{\id_E\otimes\Phi}_{=:\Psi_E}\otimes\id_{\cX}\|_{(E:q,Q:q;\cX)\to(E:q,R:q,S:p;\cX)}\\ &\equiv \sup_E\|\Psi_E\otimes\id_{\cX}\|_{(EQ:q;\cX)\to(ER:q,S:p;\cX)} \\&{=} \sup_E\|\Psi_E\|^+_{(EQ:q)\to(ER:q,S:p)}=\sup_E\|\id_E\otimes\Phi\|^+_{(E:q,Q:q)\to(E:q,R:q,S:p)}\\&= \|\Phi\|^+_{cb,(Q:q)\to(R:q,S:p)}.  
    \end{align}
\end{proof}
This result will be important in proving  \cref{thm:mainchainrule}. It also has an interpretation as a chain rule for R\'{e}nyi-entropies when fixing $q=1$. Like above applying $\frac{\alpha}{1-\alpha}\log$ to the above directly yields
\begin{corollary}\label{cor:chain.rule.cb.version}
Let $\Phi:Q\to RS$ be a CP map
For any system $E$ and any state $\rho\in \mathcal{D}(\cH_{E}\otimes\cH_{Q} \otimes \cH_{T})$, we have the following chain rule 
\begin{align}
H^{\uparrow}_\alpha(ST|RE)_{(\id_E\otimes\Phi 
\otimes \id_T)(\rho_{EQT})} - H^{\uparrow}_\alpha(T|QE)_\rho &\geq   \inf_E\inf_{\sigma \in \mathcal{D}(\cH_{QE})} H^{\uparrow}_\alpha(S|RE)_{(\id_E\otimes\Phi)(\sigma_{EQ})}\,.
\end{align}  
\end{corollary}

\subsection{Additivity}

Making use of the previously established technical Lemmas, we present a general multiplicativity result for CB norms:
\begin{theorem}[Multiplicativity of ordered CB norms]\label{thm:general.multiplicativity}
Let $\cX, \cY$ be operator spaces and $1 \leq p,q \leq \infty$. Let $\Phi:\cS_q(\mathcal{H}_Q)\to \cS_p(\mathcal{H}_P),\Psi:\mathcal{X}\to \mathcal{Y}$ be CP maps, then
\begin{align}
\|\Phi\otimes\Psi\|_{cb,S_q[\mathcal{H}_Q,\mathcal{X}]\to S_p[\mathcal{H}_P,\mathcal{Y}]} = \|\Phi\|_{cb,(Q:q)\to (P:p)}\|\Psi\|_{cb,\mathcal{X}\to \mathcal{Y}}.
\end{align} And as a direct consequence, it holds for CP maps $\{\Phi_i:Q_i\to P_i\}$ and numbers $1\leq q_i,p_i\leq\infty$ that
\begin{align}
    \|\bigotimes_{i=1}^n\Phi_i\|_{cb,(Q_1:q_1,...,Q_n:q_n)\to (P_1:p_1,...,P_n:p_n)} = \prod_{i=1}^n\|\Phi_i\|^+_{cb,(Q_i:q_i)\to (P_i:p_i)}.
\end{align}
\end{theorem}

\begin{remark}
    This result is a generalization of \cite[Theorem 11 and Theorem 13 c)]{Devetak.2006} that corresponds to the case $q=q_1 = \cdots = q_n$ and $p=p_1 = \cdots = p_n$. Note that such multiplicativity statements under tensor products do not hold in general for non-CB norm, see e.g., \cite[Section 5]{Devetak.2006}. 
\end{remark}
\begin{proof}[Proof of \cref{thm:general.multiplicativity}]
This proof of the upper bound follows from a combination of \cref{lem:cb.norm.splitting} and \cref{lem:reductionLemma}. 
We apply the former to the maps $\Phi\otimes\id_{\mathcal{Y}}:\mathcal{S}_q[\mathcal{H}_Q,\mathcal{Y}]\to \mathcal{S}_p[\mathcal{H_P,\mathcal{Y}}]$ and $(\id_Q\otimes\Psi):\mathcal{S}_q[\mathcal{H}_Q,\mathcal{X}]\to \mathcal{S}_q[\mathcal{H}_Q,\mathcal{Y}]$ to get
\begin{align}
\|\Phi\otimes\Psi\|_{cb,S_q[\mathcal{H}_Q,\mathcal{X}]\to S_p[\mathcal{H}_P,\mathcal{Y}]} &\leq \|\Phi\otimes\id_\mathcal{Y}\|_{cb,\mathcal{S}_q[\mathcal{H}_Q,\mathcal{Y}]\to \mathcal{S}_p[\mathcal{H_P,\mathcal{Y}}]} \cdot \|\id_Q\otimes\Psi\|_{cb,\mathcal{S}_q[\mathcal{H}_Q,\mathcal{X}]\to \mathcal{S}_q[\mathcal{H_Q,\mathcal{Y}}]} \\ &= 
\|\Phi\|^+_{cb,Q:q\to P:p}\cdot \|\Psi\|_{cb,\mathcal{X}\to\mathcal{Y}},
\end{align} where the last line follows from \cref{lem:reductionLemma} and the absorption of the $\id_Q$ on the left is due to the definition of the CB norm.
For the other inequality, for some system $E_1, E_2$, let $X_{E_1Q} \in \cS_p[\cH_{E_1}, \cS_q(\cH_{Q})]$ and $Y_{E_2 \cX} \in \cS_{p}[\cH_{E_2}, \cX]$. 

Let $E=E_1E_2$, then
\begin{align}
\|\Phi\otimes\Psi\|_{cb,S_q[\mathcal{H}_Q,\mathcal{X}]\to S_p[\mathcal{H}_P,\mathcal{Y}]} &\geq \|(\id_{E_1}\otimes\id_{E_2}\otimes\Phi\otimes\Psi)(X_{E_1Q}\otimes Y_{E_2\mathcal{X}})\|_{(E_1:p,E_2:p,P:p;\mathcal{Y})} \\
    &= \|(\id_{E_1}\otimes\Phi)(X_{E_1Q})\otimes(\id_{E_2}\otimes\Psi)( Y_{E_2\mathcal{X}})\|_{(E_1:p,P:p,E_2:p;\mathcal{Y})} \\&= \|(\id_{E_1}\otimes\Phi)(X_{EQ})\|_{(E_1:p,P:p)}\cdot \|(\id_{E_2}\otimes\Psi)( Y_{E_2\mathcal{X}})\|_{(E_2:p;\mathcal{Y})},
\end{align} where the last equality follows from \cref{prop:spliting.systems.left.} applied to $E_1P,E_2\mathcal{Y}$. Now taking the supremum over $X,Y$ and $E_1,E_2$ yields the claim.

The multiplicativity result for $n$ tensored CP maps follows now directly via induction. For simplicity denote with $Q^n_j:=Q_j...Q_n$ and with $q_j^n:=(q_j,...,q_n)$ and similarly for $P, p$. Now the above is the induction start and the step follows via
\begin{align}
\bigg\|\bigotimes_{i=1}^n\Phi_i\bigg\|_{cb,(Q^n:q^n)\to (P^n:p^n)} &= \bigg\|\Phi_1\otimes\bigotimes_{i=2}^n\Phi_i\bigg\|_{cb,(Q_1:q_1,Q_2^n:q_2^n)\to (P_1:p_1,P_2^n:p_2^n)} \\ 
    &=\|\Phi_i\|_{cb,(Q_1:q_1)\to (P_1:p_1)}\cdot \bigg\|\bigotimes_{i=2}^n\Phi_i\bigg\|_{cb,(Q_2^n:q_2^n)\to (P_2^n:p_2^n)},
\end{align} where in the second line we used the above with $\mathcal{X}=\mathcal{S}_{q_2}[...\mathcal{S}_{q_n}(\mathcal{H}_{Q_n})...]$ and $\mathcal{Y}=\mathcal{S}_{p_2}[...\mathcal{S}_{p_n}(\mathcal{H}_{P_n})...]$.
\end{proof}

As a direct consequence, we get a special case of the generalized EAT chain rule~\cite{Metger.2024} for product quantum channels.

\begin{corollary}[A chain rule under product maps]\label{cor:generalized.eat.chain.rule}
Consider a CP map of product form $\Phi_{Q_1Q_2 \to RS} = \phi_{Q_1 \to R} \otimes \psi_{Q_2 \to S}$. Then we have
\begin{align}
\label{equ:Generalized.EAT.product.norm}
\|\Phi\otimes\id_T\|_{(Q_1:1,T:\alpha,Q_2:1)\to(R:1,S:\alpha,T:\alpha)} \leq \|\Phi\|_{cb,(Q_1:1,Q_2:1)\to (R:1,S:\alpha)} 
\end{align}
which implies
\begin{align}\label{equ:Generalized.EAT.product}
H^{\uparrow}_\alpha(ST|R)_{(\Phi 
\otimes \id_T)(\rho_{Q_{1}Q_{2}T})} \geq H^{\uparrow}_\alpha(T|Q_1)_{\rho} + \inf_{\sigma \in \mathcal{D}(\cH_{Q_1Q_2} \otimes \cH_{\tilde{Q}})} H^{\uparrow}_\alpha(S|R \tilde{Q})_{(\Phi \otimes \id_{\tilde{Q}})(\sigma_{Q_1Q_2\tilde{Q}})},
\end{align} where $\tilde{Q}$ is a purifying system isomorphic to $Q$. 
\end{corollary}
\begin{remark}
This chain rule~\eqref{equ:Generalized.EAT.product} is similar to the one in~\cite[Lemma 3.6]{Metger.2024} with the replacements $Q_1 \to E$, $Q_2 \to R$, $S \to A'$, $T \to A$, $R \to E'$. The differences are the we assume the product condition which is stronger than the non-signalling condition in~\cite[Lemma 3.6]{Metger.2024}, but we use $H_{\alpha}^{\uparrow}$ instead of $H_{\alpha}$ and our chain rule is applicable to any $\alpha \geq 1$ and there is no loss in the parameter $\alpha$. 
\end{remark}
\begin{proof}
We first establish~\eqref{equ:Generalized.EAT.product.norm}. By the fact that we can combine systems with the same parameter (\cref{prop:combining.systems}) and that the completely bounded norm is multiplicative (\cref{thm:general.multiplicativity}), we have
\begin{align}
\| \Phi \otimes \id_{T} \|_{(Q_1:1,T:\alpha,Q_2:1)\to(R:1,S:\alpha,T:\alpha)} 
&= \|\phi_{Q_1\to R}\otimes\id_T\otimes\psi_{Q_2\to S}\|_{(Q_1:1,T:\alpha,Q_2:1)\to(R:1,S:\alpha,T:\alpha)} \\
&= \|\phi_{Q_1\to R}\otimes\id_T\otimes\psi_{Q_2\to S}\|_{(Q_1:1,T:\alpha,Q_2:1)\to(R:1,T:\alpha,S:\alpha)} \\
&\leq \|\phi_{Q_1\to R}\otimes\id_T\otimes\psi_{Q_2\to S}\|_{cb,(Q_1:1,T:\alpha,Q_2:1)\to(R:1,T:\alpha,S:\alpha)} \\
&= \|\phi\|_{cb,(Q_1:1) \to (R:1)} \|\id_T\|_{cb,T:\alpha \to T:\alpha} \|\psi_{Q_2\to S}\|_{cb,(Q_2:1) \to (S:\alpha)} \\
&= \|\phi \otimes \psi\|_{cb,(Q_1:1,Q_2:1) \to (R:1,S:\alpha)} \\
&= \| \Phi \|_{cb,(Q_1:1,Q_2:1) \to (R:1,S:\alpha)}
= \| \Phi \|_{cb,(Q_1Q_2:1) \to (R:1,S:\alpha)}^+\,.
\end{align}
We now show how to deduce the chain rule~\eqref{equ:Generalized.EAT.product}. As before, $\frac{\alpha}{1-\alpha} \log \| \Phi \|_{cb,(Q_1:1,Q_2:1) \to (R:1,S:\alpha)} = \inf_{\sigma \in \mathcal{D}(\cH_{Q_1Q_2} \otimes \cH_{\tilde{Q}})} H_{\alpha}^{\uparrow}(S|R\tilde{Q})$. Moreover, given a positive semidefinite matrix $\rho_{Q_1Q_2T}$, we have that $H_{\alpha}^{\uparrow}(T|Q_1)_{\rho} = \frac{\alpha}{1-\alpha} \log \| \rho \|_{(Q_1:1,T:\alpha, Q_2:1)}$ because $\| \rho \|_{(Q_1:1,T:\alpha, Q_2:1)} = \|\tr_{Q_2} \rho_{Q_1TQ_2} \|_{(Q_1:1,T:\alpha)}$ as proved in~\cite[Section 3.5]{Devetak.2006}. Furthermore, $H_{\alpha}^{\uparrow}(ST|R)_{(\Phi \otimes \id_T)(\rho)} = \frac{\alpha}{1-\alpha} \log \|(\Phi \otimes \id_T)(\rho)\|_{(R:1,S:\alpha,T:\alpha)}$.
\end{proof}

Motivated by applications in~ \cref{sec:Applications.QKD}, we now consider multiplicativity with a different order. The maps $\Phi$ and $\Psi$ have composite output systems $R_1S_1$ and $R_2S_2$ and the relevant norm on the output is a multi-index Schatten norm in the order $R_1R_2S_1S_2$. We also restrict ourselves to an index $q$ for the $R$ systems (the index of the input systems $Q$) and an index $p$ for the $S$ systems.


\begin{theorem}[Multiplicativity of $q \to (q,p)-$CB-norms] \label{thm:mainchainrule} 
Let $1 \leq q,p \leq \infty$. Let $\Phi:Q_1\to R_1S_1$ and $\Psi:Q_2\to R_2S_2$ be two CP maps, then writing $Q^2:=Q_1Q_2$, $R^2=R_1R_2$, and $S^2=S_1S_2$, it holds that
    \begin{align}
        \|\Phi\otimes\Psi\|_{cb,(Q^2:q)\to (R^2:q,S^2:p)} \leq \|\Phi\|^{+}_{cb,(Q_1:q)\to({R_1}:q,{S_1}:p)} \cdot \|\Psi\|^{+}_{cb,(Q_2:q)\to(R_2:q,{S_2}:p)}. 
    \end{align} 
       As a direct consequence, it holds for CP maps $\{\Phi_i:Q_i\to R_iS_i\}$ that
    \begin{align}
   \bigg\|\bigotimes^n_{i=1}\Phi_i\bigg\|_{cb,(Q^n:q)\to (R^n:q,S^n:p)}\leq \prod_{i=1}^n\|\Phi_i\|_{cb,{Q_i}:q\to ({R_i}:q, {S_i}:p)},
    \end{align}
    where we denoted $Q^n:=Q_1...Q_n$, $R^n:=R_1...R_n, S^n:=S_1...S_n$.
\end{theorem}   
\begin{remark}
Note firstly that this result is also a generalization of \cite[Theorem 11 and Theorem 13 c)]{Devetak.2006} but in a different way: we recover the upper bounds of \cite[Theorem 11 and Theorem 13 c)]{Devetak.2006} by letting the systems $R$ be trivial. Secondly, note that due to \cref{lem:positivesufficiency} and \cref{Cor:rightIdentity} the optimization in the CB norms on both sides can be restricted to positive states only.    
\end{remark}

\begin{proof}[Proof of \cref{thm:mainchainrule}]
Note that we may assume that all the norms are finite, otherwise the equality clearly holds.
We apply \cref{lem:cb.norm.splitting} and write $\Phi \otimes \Psi$ as a composition of four maps, since the order in the multi-index norm we are considering does not respect the tensor product structure of the maps: $\Phi \otimes \Psi.$ We write it as $(F_{R_1\leftrightarrow R_2}\otimes\id_{S_1S_2})\circ(\id_{R_2} \otimes \Phi \otimes \id_{S_2}) \circ (F_{Q_1 \leftrightarrow R_2} \otimes \id_{S_2}) \circ (\id_{Q_1} \otimes \Psi)$. Note that we wrote the swap explicitly to emphasize the order in the multi-index Schatten norms that we use. For additional clarity, we specify the operator spaces for each map: The input operator space for $(\id_{Q_1} \otimes \Psi)$ is $\cX_1 = \cS_q[\cH_{Q_1},\cS_q(\cH_{Q_2})]$ and the output is $\cX_2 = \cS_q[\cH_{Q_1}, \cS_{q}[\cH_{R_2}, \cS_{p}(\cH_{S_2})]]$. The output operator space of $(F_{Q_1 \leftrightarrow R_2} \otimes \id_{S_2})$ is $\cX_3 = \cS_q[\cH_{R_2}, \cS_{q}[\cH_{Q_1}, \cS_{p}(\cH_{S_2})]]$ and the output operator space of $(\id_{R_2} \otimes \Phi \otimes \id_{S_2})$ is $\cX_4 = \cS_q[\cH_{R_2}, \cS_{q}[\cH_{R_1}, \cS_{p}[\cH_{S_1}, \cS_{p}(\cH_{S_2})]]]$. The last map now maps $\mathcal{X}_4$ into $\mathcal{X}_5:=\cS_q[\cH_{R_1}, \cS_{q}[\cH_{R_2}, \cS_{p}[\cH_{S_1}, \cS_{p}(\cH_{S_2})]]]$. It now remains to bound the CB norm of each one of these three maps. First, by definition of the CB norm, we have
\[
\| (\id_{Q_1} \otimes \Psi) \|_{cb,\cX_1 \to \cX_2} = \| \Psi \|_{cb, \cS_q(\cH_{Q_2}) \to \cS_{q}[\cH_{R_2}, \cS_{p}(\cH_{S_2})]}.
\]
Second, using the fact that the swap between systems of equal Schatten indices is a complete isometry and its CB norm is not affected by an identity on the right~\eqref{equ:SWAP.isometry} gives
\begin{align}
 \|F_{Q_1 \leftrightarrow R_2} \otimes \id_{S_2} \|_{cb, \cX_2 \to \cX_3} &= 1, \\
 \|F_{R_1\leftrightarrow R_2}\otimes\id_{S_1S_2}\|_{cb,\mathcal{X}_4\to\mathcal{X}_5} &= 1.
\end{align}
Third, using \cref{Cor:rightIdentity} and also the definition of the completely bounded norm
\[
\| \id_{R_2} \otimes \Phi \otimes \id_{S_2} \|_{cb,\cX_3 \to \cX_4} = \| \Phi \|_{cb,\cS_q(\cH_{Q_1}) \to \cS_{q}[\cH_{R_1},\cS_p(\cH_{S_1})]}.
\]
Finally, we get
\begin{align} \label{equ:inductionStep}
\|\Phi\otimes\Psi\|_{cb,(Q_1:q,Q_2:q)\to(R_1:q,R_2:q,S_1:p,S_2:p)} 
\leq \|\Phi\|_{cb,(Q_1:q)\to(R_1:q,S_1:p)} \cdot \|\Psi\|_{cb,(Q_2:q)\to(R_2:q,S_2:p)}.
\end{align} 

The proof of the $n$-fold statement follows similarly to in \cref{thm:general.multiplicativity} by induction over the multiplicativity statement.
Denote with $Q^j:=Q_1...Q_j$ for $1\leq j\leq n$, and analogously $R^j,S^j$. The theorem is the induction start, for the induction step observe
\begin{align}
    \bigg\|\bigotimes^n_{i=1}\Phi_i\bigg\|_{cb,(Q:q)\to (R:q,S:p)} &= \left\|\bigotimes^{n-1}_{i=1}\Phi_i\otimes\Phi_n\right\|_{cb,({Q^{n-1}:q},{Q_n}:q)\to (R^{n-1}:q,{R_n}:q,S^{n-1}:p,S_n:p)} \\ &\leq \bigg\|\bigotimes^{n-1}_{i=1}\Phi_i\bigg\|_{cb,(Q^{n-1}:q)\to (R^{n-1}:q,S^{n-1}:p)}\cdot\big\|\Phi_n\big\|_{cb,(Q_n:q)\to (R_n:q,S_n:p)}, 
\end{align} where the inequality in the second line follows from \eqref{equ:inductionStep}.
\end{proof}

We also obtain a submultiplicativity result for the sequential composition of CP maps.
\begin{theorem}[Sequential composition of maps]\label{thm:Sequential.chain.rule}
Let $\Phi:Q_1\to R_1Q_2S_1, \Psi:Q_2\to R_2S_2$ be two CP maps and write $R^2:=R_1R_2$, $S^2:=S_1S_2$, then for any $1\leq q,r,s\leq\infty$
    \begin{align}
\|\Psi\circ\Phi\|_{cb,(Q_1:q)\to(R^2:r,S^2:s)} &\leq \|\Psi\|^{+}_{cb,(Q_2:q)\to(R_2:r,S_2:s)} \cdot \|\Phi\|^{+}_{cb,(Q_1:q)\to(R_1:r,Q_2:q,S_1:s)}.
    \end{align} 
\end{theorem}

\begin{proof}
Using the fact that consecutive systems with the same Schatten index can be swapped (\cref{prop:combining.systems}), we can write $\|\Psi\circ\Phi\|_{cb,(Q_1:q)\to(R^2:r,S^2:s)} = \|\Psi\circ\Phi\|_{cb,(Q_1:q)\to(R_1:r,R_2:r,S_2:s,S_1:s)}$. We then write the composition more explicitly including identities: $\Psi\circ\Phi =  (\id_{R_1} \otimes \Psi \otimes \id_{S_1}) \circ \Phi$ and then use~\cref{lem:cb.norm.splitting} to get
\begin{align}
&\|\Psi\circ\Phi\|_{cb,(Q_1:q)\to(R_1:r,R_2:r,S^2:s,S_1:s)} \\
&\leq \| \id_{R_1} \otimes \Psi \otimes \id_{S_1} \|_{cb,(R_1:r,Q_2:q,S_1:s) \to (R_1:r,R_2:r,S_2:s,S_1:s)} \| \Phi \|_{cb,(Q_1:q) \to (R_1:r,Q_2:q,S_1:s)}.
\end{align}
Since $\Phi$ is CP, by \cref{lem:positivesufficiency} we have $\| \Phi \|_{cb,(Q_1:q) \to (R_1:r,Q_2:q,S_1:s)} = \| \Phi \|_{cb,(Q_1:q) \to (R_1:r,Q_2:q,S_1:s)}^+$. Then by definition of the CB norm and then using \cref{Cor:rightIdentity}, we have 
\begin{align}
\| \id_{R_1} \otimes \Psi \otimes \id_{S_1} \|_{cb,(R_1:r,Q_2:q,S_1:s) \to (R_1:r,R_2:r,S_2:s,S_1:s)} 
&= \| \Psi \otimes \id_{S_1} \|_{cb,(Q_2:q,S_1:s) \to (R_2:r,S_2:s,S_1:s)} \\
&= \| \Psi \|^+_{cb,(Q_2:q) \to (R_2:r,S_2:s)}\,,
\end{align}
which proves the desired result.

\end{proof}

\subsection{Linear constraint setting and additivity}

For the applications in quantum cryptography that we present in the next section, it is important to be able to restrict the optimization with a linear constraint. So we consider \cref{thm:mainchainrule} for $q = 1$ by restricting the optimization implicit in the CB-norms to states satisfying some linear constraints. To enforce the linear constraints, we introduce the following definitions.

\begin{definition}[Restricted state spaces]
    Let $\mathcal{H}_Q$ be the Hilbert space of a quantum system. Then we specify a linear constraint on $\mathcal{D}(\mathcal{H}_Q)$ by a linear CPTP map $\mathcal{N}:Q\to Q^\prime$ and a state $\tau\in\mathcal{D}(\mathcal{H}_{Q^\prime})$, where $Q^\prime$ is some other quantum system. 
Given such a tuple $r=(\mathcal{N},\tau)$ we define
\begin{equation}\label{def:reduced.state.spaces}
    \mathcal{D}_r(\cH_Q) := \{\rho \in \Pos(Q)|\mathcal N(\rho) = \tau\Tr[\rho]\}.
    \end{equation}    
More generally, let $\{\mathcal{D}_{r_i}(\cH_{Q_i})\}_{i\leq n}$ be $n$ such restricted sets of states defined in terms of the tuples $r^n=\{r_i\}_{i=1}^n=\{(\mathcal N_i,\tau_i)\}^n_{i=1}$ and $E$ any other quantum system. Then we define
\begin{equation}\label{eq:reduced.state.spaces.with.Environement}
    \mathcal{D}^E_{r^n}(\cH_{Q^n})= \left\{ \rho\in \Pos(EQ^n)|\textstyle\left(\bigotimes_{i\leq n}\mathcal N_i\right) \circ \tr_E[\rho] = \textstyle\bigotimes_{i\leq n}\tau_i\Tr[\rho]\right\}.
\end{equation}
\end{definition}
In particular, when $\mathcal{N}:Q\to \mathbb{C}, \rho\mapsto\Tr[\rho]$ is the full trace and $\tau = 1$, then $\mathcal{D}_r(\cH_Q) = \mathcal{D}(\cH_Q)$ and $\mathcal{D}^E_r(\cH_Q) = \mathcal{D}(\cH_{EQ})$ is just the set of unrestricted normalized states. The restricted CB-norms of quantum channels are then naturally defined as follows.
\begin{definition}[Restricted CB-norms]
Let $r:=(\mathcal{N},\tau)$ define a linear restriction and $\Phi:Q\to RS$ be a CP map, then we define the \textit{restricted completely bounded norm} of $\Phi$ as
\begin{align}
\|\Phi\|_{r,cb,(Q:q)\to(R:r,S:s)}:=\sup_E\sup_{\rho\in \mathcal{D}_r^E(\cH_Q)}\frac{\|(\id_E\otimes\Phi)(\rho)\|_{(E:q,R:r,S:s)}}{\|\rho\|_{(E:q,Q:q)}}.
\end{align}
\end{definition}

Now, we may strengthen \cref{thm:mainchainrule} to also hold under arbitrary linear constraints of the form~\eqref{eq:reduced.state.spaces.with.Environement}. 

\begin{theorem}[Multiplicativity of restricted CB $1\to (1,p)$ norms]\label{thm:Restricted.multiplicativity}
Let $n$ CP maps $\Phi_i:Q_i\to R_iS_i$ such that $\|\Phi_i\|_{cb,(Q_i:1)\to(R_i:1,S_i:p)}<\infty$ and $n$ linear restrictions governed by triples $r_i:=(Q_i^\prime,\mathcal{N}_i,\tau_i)$ be given. Denote the combined spaces $Q^n:=Q_1...Q_n$, $R^n:=R_1...R_n$, and $S^n:=S_1...S_n$ and the combined linear restriction as $r^n:=(\otimes_{i=1}^n\mathcal{N}_i,\otimes_{i=1}^n\tau_i)$.
Then it holds that
    \begin{align}
        \label{eq:multiplicativity.1to1P}
        \bigg\| \bigotimes_{i=1}^n \Phi_i\bigg\|_{r^n,cb,(Q^n:1)\to (R^n:1,S^n:p)} 
            &\leq \prod^n_{i=1} \| \Phi_i\|_{r_i,cb,(Q_i:1)\to (R_i:1,S_i:p)}\,.
    \end{align}
\end{theorem}
\begin{remark} This is a generalization of a result from \cite{Himbeeck.2025}, in which this statement, formulated in terms of Rényi entropies and derived via very different tools, was shown to hold in the special case where all channels $\Phi_i=\mathcal{M}$ are equal.
\end{remark}
\begin{proof} 
The proof follows via lifting \cref{thm:mainchainrule} to the restricted setting, in the same manner as was done in \cite{Himbeeck.2025}. Hence it is repeated in \cref{app:Additivity.on.reduced.operator.spaces}.
\end{proof}


\section{Applications to quantum cryptography}
\label{sec:Applications.QKD}

We now show how the multiplicativity of restricted CB $1\to (1,p)$ norms shown in Theorem~\ref{thm:Restricted.multiplicativity} can be used to prove the security of time-adaptive quantum cryptographic protocols. In this section we will assume all Hilbert spaces to be finite dimensional.

\subsection{$f$-weighted Rényi entropy}

A key quantity used in security proofs is the $f$-weighted Rényi entropy \cite{Himbeeck.2025}, which is defined as follows. Let $\mathbb X$ be a finite set, $f:\mathbb X\to \mathbb R$ a function, and $\rho_{XAE} = \sum_x \ketbra{x}_X \otimes \rho_{AE}^x $ a state that is classical on system $X$, then for all $\alpha>1$, define the $f$-weighted Renyi entropy as
\begin{align}
    H_\alpha^{\uparrow,f}(A|XE)_{\rho} 
        &:= \frac{\alpha}{1-\alpha} \log \| 2^{\tfrac{\alpha-1}{\alpha}f_X} \cdot \rho_{EXA}\|_{(EX:1,A:\alpha)}\\
        &= \frac{\alpha}{1-\alpha} \log \sum_{x\in \mathbb X} 2^{\tfrac{\alpha-1}{\alpha}f(x)} \|\rho_{EA}^x\|_{(E:1,A:\alpha)}
\end{align}
where $f_X = \sum_x f(x) \ketbra{x}$ is a diagonal operator, which commutes with $\rho_{EXA}$. In particular, we note that when $f(x)=0$ for all $x\in \mathbb X$, we recover the usual Rényi entropy $H_\alpha^{\uparrow,f}(A|XE)_{\rho} = H_\alpha^{\uparrow}(A|XE)_{\rho}$.

As we will see now, Theorem~\ref{thm:Restricted.multiplicativity} implies a chain rule for $f$-weighted Rényi entropies. Assume we need to minimize the $f$-weighted Rényi entropy over a set of states of the form
\begin{align}
    \{ \rho_{EX^nA^n} = (\id_E\otimes\mathcal{M}^n) (\rho_{EQ^n}) \: | \: \rho_{EQ^n} \in \mathcal D_r^E(\cH_{Q^n})\} 
\end{align} 
for all possible choices of environments $E$, where $\mathcal M^n = \bigotimes_{i=1}^n \mathcal M_i$ and $\mathcal M_i:Q_i\to X_iA_i$ are CP map and $\mathcal{D}^E_{r^n}(\cH_{Q^n})$ is an environment embedded restricted state space as defined in \eqref{eq:reduced.state.spaces.with.Environement}. Assume moreover that $f^n(x^n)= \sum_i f_i(x_i)$. Then the results below shows that the minimium is obtained for an tensor product state of the form $\rho_{{E^\prime}^nX^nA^n} = \bigotimes_{i=1}^n \rho_{E^\prime XA}^i$ with $\rho_{E^\prime XA}^i =  \id_{E^\prime}\otimes\mathcal{M}_i(\rho^i_{E^\prime Q})$ and $\rho_{QE'}^i\in  D_r^{E'}(\cH_{Q_i})$ and $E = {E'}^n$, in which case then entropy is additive. 

\begin{corollary}[Reduction to independent attacks]
    \label{thm:IIDreduction}
    For all $i\in [n]$, let $\mathcal M_i:Q_i\to X_iA_i$ be a CP map between finite dimensional systems, with $X_i$ a classical register with basis elements labeled $\mathbb{X}$, and let $f_i : \mathbb{X} \to \mathbb{R}$. Set $Q^n:=Q_1...Q_n, X^n:=X_1...X_n,$ and $A^n:=A_1...A_n$ and define $\mathcal M^n = \bigotimes_{i=1}^n \mathcal M_i$ and  $f^n(x^n) = \sum_{i=1}^n f_i(x_i)$. Then for all $\alpha>1$
    \begin{align}
        \inf_E\inf_{\rho\in \mathcal{D}^E_r(\cH_{Q^n})} H_\alpha^{\uparrow,f_n}(A^n|X^nE)_{(\id_E\otimes \mathcal M^n)(\rho)} 
            &= \sum_{i=1}^n \inf_E\inf_{\rho_i\in \mathcal{D}_r^E(\cH_{Q_i})} H_\alpha^{\uparrow,f_i}(A_i|X_iE)_{(\id_E\otimes\mathcal M_i)(\rho_i)} \\ 
    \end{align} 
\end{corollary}
\begin{proof}
Define the operators $f_{i,X_i} := \sum_{x \in \mathbb{X}} f_i(x) \ketbra{x}_{X_i}$ and $f^n_{X^n} = \sum_{x^n\in \mathbb{X}^n} f^n(x^n)\ketbra{x^n}_{X^n}$ and the maps $\Phi_i : Q_i \to X_i A_i $ defined by $\Phi_i(\rho) := 2^{\tfrac{\alpha-1}{2\alpha}f_{i,X_i}} \mathcal{M}_i(\rho) 2^{\tfrac{\alpha-1}{2\alpha}f_{i,X_i}}$, which are CP by construction, since $2^{\frac{\alpha-1}{2\alpha}f_{i,X_i}}$ are self-adjoint. Then, we can use Theorem~\ref{thm:Restricted.multiplicativity} for the maps $\Phi_i$ with the replacements $A_i \to S_i$ and $X_i \to R_i$  and by applying $\frac{\alpha}{1-\alpha}\log$ to each side of the equation \eqref{eq:multiplicativity.1to1P} and noting that $\bigotimes_{i=1}^n 2^{f_{i,X_i}} = 2^{f^n_{X^n}}$ we get that the LHS in upper bounded by the RHS.
Equality follows from additivity of $H^\uparrow_\alpha$ under tensor products of states \cite[Corollary 5.9]{Book.Tomamichel.2016}.
\end{proof}

\subsection{Definition of the protocol}
\label{subsec:protocol_definition}
For simplification, we will consider a random number generation (QRNG) protocol, which are closely related to quantum key distribution (QKD) protocols. Using standard techniques, the present security proof can be easily generalized to QKD \cite{Renner.2006, Himbeeck.2025}. 

An $n$-round device-dependent random number generation protocol consists of two steps. First, we generate the raw data and then determine how much secure randomness can be extracted from that data. The first step can be represented as a CPTP map
\begin{align}
    \mathcal M^n: Q^n \to X^nA^n
\end{align}
from an input space $Q^n=Q_1 \dots Q_n$ to the output systems $X^n$ and $A^n$, representing respectively the public announcements and the raw key. Since they are classical variables, 
we can write operators on the corresponding Hilbert spaces as diagonal operators in some canonical basis, the elements of which are labeled by the finite sets $\mathbb{X},\mathbb{A}$. 


\paragraph{Independent rounds}
We assume that our protocol varies over times, but that different rounds of the protocol, corresponding to the CPTP maps $\mathcal M_t: \cD(\cH_{Q})\to \cD(\cH_{XA})$, act independently on different inputs, i.e.
\begin{align}
    \mathcal M^n = \bigotimes_{t=1}^n \mathcal M_t\,.
\end{align}

\paragraph{Linear constraint on the input}
We assume that the protocol is applied to an initial unknown input state $\rho_{EQ^n}$ entangled with an arbitrary reference system $E$, and where $\rho_{Q^n} = \Tr_E[\rho_{EQ^n}]$ satisfies a linear constraint of the form 
\begin{align}
    \label{eq:assumptionQKD}
    \bigotimes_t \mathcal{N}_t(\rho_{Q^n}) = \bigotimes_t \tau_t
\end{align}
where $\cN_t: \cD(\cH_Q)\to \cD(\cH_{Q'})$ is a completely positive map to some system $Q'$ and $\tau_t\geq 0$ a positive semidefinite operator on $Q'$. This last condition is used in prepare-and-measure quantum key distribution protocols \cite{Winick.2018}. This corresponds to saying that $\rho_{EQ^n}\in \mathcal D^E_{r^n}(\cH_{Q^n})$ for some restricted state space defined in \eqref{eq:reduced.state.spaces.with.Environement}.

\subsection{Security and rate of the QRNG protocol}
\paragraph{Post-processing}
To complete the protocol, we need to specify how we determine the amount of randomness that can be extracted from the raw key registers $A^n$. This is done using some function $g_n:\mathbb{X}^n\to \mathbb N$, which will be constructed below. The protocol first evaluates $k \leftarrow g_n(x^n)$, then samples an extractor $E_s: \mathbb{A}^n \to \{0,1\}^k$ from a 2-universal family with seed $s$, and finally applies the extractor to the raw key register $A^n$ and writes the $k$-bit result in the classical register $K_{x^n}$ holding bitstrings of length $g_n(x^n)$.
We write the map performing this as $\mathcal{R}^{g_n}: \cD(\cH_{X^nA^n}) \to \cD(\oplus_{x^n} \cH_{K_{x^n}} \otimes \cH_{S} )$ for this map.

Note that the length of the key is itself a random variable, whose distribution depends on the input state $\rho_{EQ^n}$. Moreover, values of $x^n\in \mathbb{X}^n$ for which $g_n(x^n)=0$, correspond to the cases where the protocol aborts.

\paragraph{Composable security}
For a given protocol $\mathcal{M}^n$, post-processing $g_n$ and input state $\rho_{Q^nE}$ satisfying \eqref{eq:assumptionQKD}, let $\oplus_{x^n \in \mathbb{X}^n} \rho_{K_{x^n}SE}^{x^n} = (\mathcal R^{g_n}\circ \cM^n \otimes \id_E)(\rho_{Q^nE})$ be the state obtained after applying the protocol and the post-processing. 
It is shown in \cite{Portmann.2022} that the composable security level is given by
\begin{align}
    \label{eq:security}
    \epsilon(\mathcal M^n,g_n,\rho) := \frac{1}{2} \sum_{x^n} \left\| \rho^{x^n}_{K_{x^n}SE} - \tfrac{\id_{K_{x^n}}}{|K_{x^n}|}\otimes \rho_{SE}^x \right\|_1\,.
\end{align}
The full protocol is said to be $\epsilon$-secure if for any input state $\rho_{Q^nE}$ satisfying \eqref{eq:assumptionQKD}, the final state
satisfies $\epsilon(\mathcal P,g_n,\rho)\leq \epsilon$.

\paragraph{Asymptotic rate} A protocol not only needs to be secure, it must also be efficient. Contray to the security condition, which must hold for any input state, the efficiency of the protocol is evaluated with respect to some ``honest" input state, which is known in advance. Assume we are given honest input states $\rho_{Q_t}^{\mathrm{hon}}$ for $t\in \mathbb N$, we consider the corresponding distributions $q^{\mathrm{hon}}_{X_t} = \tr_{A} \circ \mathcal M_t(\rho_{Q_t}^{\mathrm{hon}})$ and $q^{\mathrm{hon}}_{X^n} = \otimes_t q^{\mathrm{hon}}_{X_t}$. Then the average key rate is given by
\begin{align}
    \mathrm{rate}(\mathcal M^n, g_n, q^{\mathrm{hon}}_{X^n}) = \frac{1}{n} \mathbb E_{x^n \sim q^{\mathrm{hon}}_{X^n}}[g_n(x^n)]\,.
\end{align}
In particular we will be interested in the asymptotic limit $n\to \infty$.

\subsection{Time adaptive asymptotic key rates}

For every $n\in \mathbb N$, we want to build a post-processing function $g_n:\mathbb{X}^n \to \mathbb R$, so that (a) the protocol is $\epsilon$-secure for any input state and (b) it achieves the largest possible key rate on average when applied to the state $\otimes_{t=1}^n \rho_{Q_t}^{\mathrm{hon}}$. How large can the average key rate be in the asymptotic limit $n\to \infty$? 

\paragraph{Time-invariant protocols} This question was considered by Renner in \cite{Renner.2006}, who considered the case where protocols do not vary in time, i.e.\ 
$\mathcal M_t = \mathcal M$, $\mathcal N_t=\mathcal N$, $\tau_i = \tau$, $q^{\mathrm{hon}}_{X_t} = q^{\mathrm{hon}}_{X} $. Then it is possible to achieve an asymptotic key rate given by the conditional von Neumann entropy, minimized over all possible states that reproduce the statistics $q_X^{\mathrm{hon}}$:
\begin{IEEEeqnarray*}{Rl'l}
    \IEEEyesnumber
    \label{eq:as_rate}
    h(\mathcal M,\mathcal N,\tau,q^{\mathrm{hon}}_X) =& \inf_{E,\rho_{QE} \in \cD(\cH_{QE})} & H(A|XE)_{\mathcal M(\rho_{QE})}\\
    & \mathrm{subject\ to} & \mathcal N(\rho_{Q}) = \tau\\
        &                   & \tr_A \circ \mathcal M(\rho_Q) = q_X^{\mathrm{hon}}\,.
\end{IEEEeqnarray*}

In other words, for every $n$ there exists a function $g_n:\mathbb X^n\to \mathbb R$, such that
\begin{IEEEeqnarray*}{rL}
    \lim_{n\to \infty} \max_{\rho_{Q^nE}} \epsilon(\mathcal M^n,g_n,\rho_{Q^nE}) &= 0\\
    \lim_{n\to \infty} \mathrm{rate}(\mathcal M^n, g_n, (q^{\mathrm{hon}}_{X})^{\otimes n}) &=  h(\mathcal M,\mathcal N,\tau,q^{\mathrm{hon}}_X).
\end{IEEEeqnarray*}

\paragraph{Time-dependent protocols}

We generalize this to the case of protocols that vary in time, where we show that it is possible to achieve an asymptotic rate given by
    \begin{align}
        r_{ad} := \lim_{n \to \infty} \frac{1}{n} \sum_{t=1}^n h(\mathcal M_t,\mathcal N_t,\tau_t,q_{X_t}^{\mathrm{hon}}) \,.
    \end{align} 

To understand the advantage, we should compare this secret key rate with the one we obtain by applying a static security proof. First, note that when the protocol and the noise are static, there is no advantage in using a time-adaptive security proof. When both vary with time, the comparison cannot be made because static methods do not apply. However, we can easily make the comparison when the protocol is static, i.e., $\mathcal M_t = \mathcal M$, $\tau_t = \tau$, $\mathcal N_t = \mathcal N$, but the noise is not, i.e. $q_{X_t}$ varies with $t$.  
    
In this case, standard proof techniques \cite{Renner.2006, Metger.2024,Himbeeck.2025} allow one to attain the secret key rate that corresponds to the average noise distribution $\bar q^{\mathrm{hon}}_X = \lim_{n\to \infty}\frac{1}{n} \sum_{t=1}^n q^{\mathrm{hon}}_{X_t}$. In other words, they allow us to construct $g_n$ such that 
\begin{align}
    \lim_{n\to \infty} \mathrm{rate}(\mathcal P, g_n,q_{X^n}) = h(\mathcal M,\mathcal N,\tau, \overline{q}^{\mathrm{hon}}_X) =: r_{na}\,,
\end{align}
However, the key rate obtained using our time-adaptive method is higher. This is because the function $h(\mathcal M, \cN, \tau, q_X)$ is a convex function in $q_X$ in general \cite{Winick.2018} and strictly convex in most cases. In the latter case, there exists distribution $q_{X_t}^{\mathrm{hon}}$ such that
\begin{align}
    r_{na} = h(\mathcal M,\mathcal N,\tau, \overline{q}^{\mathrm{hon}}_X) < \frac{1}{n} \sum_{t=1}^n h(\mathcal M,\mathcal N,\tau,q^{\mathrm{hon}}_{X_t}) = r_{ad}\,,
\end{align}
which shows that the asymptotic key rate is higher using adaptive methods. 
\subsection{Security proof}
We show that we can achieve the adaptive rate $r_{ad}$, under a technical assumptions on the honest distribution.
\begin{theorem}[Time-adaptive protocol]
    \label{thm:time_adaptive_protocol}
    Let $\mathcal{M}^n$ be a family of protocols as defined in subsection~\ref{subsec:protocol_definition} and let $\rho^{\mathrm{hon}}_{Q_t} \in \cD(\cH_{Q_t})$ for $t \in \mathbb{N}$ be a family of quantum states such that $\{(\mathcal M_t,\mathcal N_t,\mathcal \tau_i,q^{\mathrm{hon}}_{X_t})|t\in \mathbb N\}$ is a finite set and $\rho^{\mathrm{hon}}_{Q_t}> 0$ for all $t\in \mathbb N$. Then there exists a family of functions $g_n : \mathbb{X}^n\to \mathbb{R}$ which, for all $n$, lead to an $\epsilon_n$-secure protocol, so that $\lim_{n\to \infty} \epsilon_n = 0$ and  
    \begin{align}
        \lim_{n\to \infty} \mathrm{rate}(\mathcal P_n, g_n,q^{\mathrm{hon}}_{X^n}) = \lim_{n \to \infty} \frac{1}{n} \sum_{t=1}^n h(\mathcal M_t,\mathcal N_t,\tau_t,q_{X_t}^{\mathrm{hon}})= r_{ad}\,.
    \end{align} 
\end{theorem}

\begin{proof}[Proof of \cref{thm:time_adaptive_protocol}]
We first explain how to construct the functions $g_n$.
It is a standard results in QKD that $ h(\mathcal M,\mathcal N,\tau,q_{X})$ is a convex function in $q_X$ (see for example \cite{Winick.2018}). For each $t\in \mathbb N$, the assumption $\rho^{\mathrm{hon}}_{Q_t} > 0$ ensures that $q_{X_t}^{\mathrm{hon}}$ is a strictly feasible distribution (i.e.\ the inequality constraints in the convex optimization problem \eqref{eq:as_rate} can be satisfied with strict inequalities). Consequently, there exists a  supporting hyperplane to the graph of the function $q_X \mapsto h(\mathcal M_t,\mathcal N_t,\tau_t, q_X)$ at the point $(q^{\mathrm{hon}}_{X_t},h(\mathcal M_t,\mathcal N_t,\tau_t, q^{\mathrm{hon}}_{X_t}))$. Moreover, we can chose the hyperplanes to be the same for all $t\in \mathbb N$ such that $(\mathcal M_t,\mathcal N_t,\mathcal \tau_t,q^{\mathrm{hon}}_{X_t})$ are the same.

Let $f_t:\mathbb{X}\to \mathbb R$ be a function parametrizing the hyperplanes, so that $\sum_x f_t(x) q_{X}(x) \leq h(\mathcal M_t,\mathcal N_t,\tau_t,q_{X})$ for all probability distributions $q_X$, and $\sum_x f_t(x) q^{\mathrm{hon}}_{X_t}(x) =h(\mathcal M,\mathcal N,\tau,q^{\mathrm{hon}}_{X_t})$. Note that a supporting hyperplane is given by an affine function, but the constant term can always be absorbed in the coefficients $f_t(x)$ since we require the inequality to hold for normalized probability distributions satisfying $\sum_{x} q_{X}(x) = 1$. 
Recalling the definition of $h$, this guarantees that for every $t$ and every $\rho_{EXA} = \mathcal M_t(\rho_{EQ})$ with $\rho_{EQ}\in \cD_r^E(\cH_Q)$,
\begin{align}
    H(A|XE)_{\rho} \geq \mathbb E_{x \sim \rho_X}[f_t(x)]\,.
\end{align}
We define our post-processing function by $g_n(x^n) = \max(0, \lfloor f^n(x^n) - \delta(\epsilon,n)\rfloor)$ where
\begin{align}
    f^n(x^n) = \sum_{t=1}^n f_t(x_t)\,,
\end{align}
and $\delta(n,\epsilon)$ will be defined below.

We now have to show that this construction $g_n$ gives a correct lower-bound on the number of bits of randomness that can be extracted from $A^n$. Using the uniform continuity of $f$-weighted Rényi entropies (Lemma~\ref{lem:continutity}), we find that for all $t \in [n]$ and all states of the form $\rho_{EXA} = \cM_t(\rho_{QE})$ with $\rho_{EQ} \in \cD_r^E(\cH_{Q})$
\begin{align}
    H_\alpha^{\uparrow,f_t}(A|XE)_{\rho} \geq -(\alpha-1) (\log \eta_t)^2 \,,
\end{align}
for $\alpha \in (1,1+1/\log \eta_t)$ where $\eta_t$ only depends on the dimension of $A$ and on $\max_x f_t(x)$ and $\min_x f_t(x)$. Moreover, since $\eta_t$ takes only a finite set of values, we can bound $\eta_t \leq \eta = \max_{t\in \mathbb N} \eta_t$.  

Using~\cref{thm:IIDreduction}, this implies that 
\begin{align}
    H_\alpha^{\uparrow, f_n}(A^n|X^nE)_{\rho} \geq - n (\alpha-1) (\log \eta)^2\,.
\end{align}
for all state of the form $\rho_{EX^nA^n} = \mathcal{M}^n(\rho_{EQ^n})$ with $\rho_{EQ^n}\in \cD_r^E(\cH_{Q^n})$. Finally, using~\cite[Theorem 1]{Himbeeck.2025}, this implies that the choice $\delta(n,\epsilon) = n (\alpha-1) (\log \eta)^2 - \frac{\alpha}{\alpha-1}\log{1/\epsilon}$ leads to an $\epsilon$-secure protocol. 

We now choose $\alpha = 1 + \tfrac{1}{\sqrt{n}}$ and $\epsilon_n = \frac{1}{n}$. Then, the average rate given by our construction is
\begin{align}
    \frac{1}{n} \mathbb E_{x^n \sim \otimes_{t=1}^n q_{X_t}^{\mathrm{hon}}}[g(x^n)]
        &\geq \frac{1}{n} \sum_{t=1}^n \sum_{x} f_t(x)q^{\mathrm{hon}}_{X_t}(x) - \frac{(\log \eta)^2}{\sqrt{n}} - \frac{2}{\sqrt{n}} \log{1/\epsilon_n} -O(1)\\
        &= \frac{1}{n} \sum_{t=1}^n h(\mathcal M_t,\mathcal N_t,\tau_t,q^{\mathrm{hon}}_{X_t}) - O(\tfrac{\log n}{\sqrt{n}})
\end{align}
which yields the stated asymptotic limit.
\end{proof}

\paragraph{Application to the BB84 protocol}

For the BB84 protocol, it is known that in static cases the asymptotic key rate is given by the Shor-Preskill formula $1-2h(p)$ where $p$ is error on the channel, $h(p) = -p \log_2 p - (1-p)\log_2 (1-p)$ is the binary entropy. Using a standard source replacement scheme, the protocol can be represented in its equivalent entanglement-based representation. In this case, the input space to a round of the protocol is the joint qubit space of Alice and Bob $Q = Q_AQ_B$ with $\mathcal{H}_{Q_A} = \mathcal{H}_{Q_B} = \mathbb{C}^{2}$. 

Let us specify what $\mathcal M,\mathcal N,\tau$ are in this case. The map $\mathcal M$ corresponds to the following physical process: Alice and Bob each generate a uniformly random basis choice, measure in the corresponding $X$ or $Z$ basis and announce their basis publicly. Alice then randomly decides if the round is a test round, with some fixed probability $p_{\mathrm{test}}$, and announces this publicly. In this case, Alice and Bob announce their measurement results publicly and Alice sets $A = \perp$; otherwise Alice assigns her measurement results to $A$. The variable $X$ regroups all public announcements that were made. 

Because of source replacement scheme, the state $\rho_{Q}$ that is input in a round of the protocol must satisfy $\tr_{Q_B}[\rho_{Q}] = \frac{\1}{2}$. Therefore we have $Q' = Q_A$, $\mathcal N = \tr_{Q_B}$ and $\tau = \frac{\1}{2}$.

We consider a family of honest implementations of the form $\rho^{\mathrm{hon}}_{Q} = (1-2p)\phi^+ + \frac{p}{2} \1_Q$, with $\phi^+$ the maximally entangled state and $p$ a parameter that varies with time, to be defined later. This corresponds to an error in the $X$ and $Z$ basis of probability $\Pr[X_A\neq X_B] = \Pr[Z_A \neq Z_B] = p$. Let $q_{X}^{\mathrm{hon}}$ be the corresponding distribution over public announcements. A standard result is that the asymptotic rate of randomness generation of the BB84 protocol is given by
\begin{align}
    h(\mathcal M,\mathcal N,\tau,q_{X}^{\mathrm{hon}}) = 1 - h(p)
\end{align}
expressed in bits per key generation round.

Consider an honest distribution where $p$ varies in times, so that $p=p_1=0.001$ for one third of the rounds and $p=p_2=0.1$ for the other two thirds, so that the average distribution corresponds to $\bar p = \frac{1}{3} p_1 + \frac{2}{3} p_2\approx 0.067$. Computing the corresponding time-adaptive and non-adaptive randomness generation rates and subtracting the cost of error correction given by $h(\bar p)$, we find that the asymptotic secure key generation rates \textup{SKrate} are
\begin{align}
    \textup{SKrate} &= \frac{1}{3}(1- h(p_1)) + \frac{2}{3} (1-h(p_2)) - h(\bar p) \approx 0.329 &\qquad \text{(time-adaptive)}\\
    \textup{SKrate} &= 1- 2 h(\bar p) \approx 0.291 &\qquad \text{(not-adaptive)}
\end{align}
This corresponds to an increase of 13\% for time-adaptive methods compared to non adaptive methods.

\section*{Acknowledgments} We would like to thank Peter Brown for helpful discussions on multiple aspects of this project and in particular regarding convex optimization problems. We thank Amir Arqand and Ernest Tan for coordinating a simultaneous submission to the arXiv. We also thank Li Gao and the anonymous TQC reviewers for helpful comments. We acknowledge funding by the European Research Council (ERC Grant AlgoQIP, Agreement No. 851716), by government grants managed by the Agence Nationale de la Recherche under the Plan France 2030 with the references ANR-22-PETQ-0009 and ANR-22-CMAS-0001, by the European Union’s Horizon Europe research and innovation programme under the project ``Quantum Secure Networks Partnership" (QSNP, grant agreement No 101114043), by the Region $\hat{\text{I}}$le-de-France in the framework of DIM QuanTiP and by the project ChistEra-2023/05/Y/ST2/00005 (MoDIC). 

\noindent \emph{Data Availability:}
Data sharing is not applicable to this article as no datasets were generated or analyzed during the current study. \\
\emph{Conflict of interests:}
The authors have no competing interests to declare that are relevant to the content of this article.

\bibliographystyle{abbrv}
\bibliography{lib}

@book{Book.Tomamichel.2016,
   title={Quantum Information Processing with Finite Resources},
   ISBN={9783319218915},
   ISSN={2197-1765},
   url={http://dx.doi.org/10.1007/978-3-319-21891-5},
   DOI={10.1007/978-3-319-21891-5},
   journal={SpringerBriefs in Mathematical Physics},
   publisher={Springer International Publishing},
   author={Tomamichel, Marco},
   year={2016} }

@book{Book.Pisier.2003, place={Cambridge}, series={London Mathematical Society Lecture Note Series}, title={Introduction to Operator Space Theory}, publisher={Cambridge University Press}, author={Pisier, Gilles}, year={2003}, collection={London Mathematical Society Lecture Note Series}}

@book{Book.Hiai.2014,
  title = {Introduction to {{Matrix Analysis}} and {{Applications}}},
  author = {Hiai, Fumio and Petz, Dénes},
  date = {2014},
  series = {Universitext},
  publisher = {Springer International Publishing},
  location = {Cham},
  doi = {10.1007/978-3-319-04150-6},
  url = {https://link.springer.com/10.1007/978-3-319-04150-6},
  urldate = {2024-06-28},
  isbn = {978-3-319-04149-0 978-3-319-04150-6},
  langid = {english},
  year= 2014}

@book{Book.Pisier.1998,
     author = {Pisier, Gilles},
     title = {Non-commutative vector valued $L_p$-spaces and completely $p$-summing map},
     series = {Ast\'erisque},
     publisher = {Soci\'et\'e math\'ematique de France},
     number = {247},
     year = {1998},
     mrnumber = {1648908},
     zbl = {0937.46056},
     language = {en},
     url ={http://www.numdam.org/item/AST_1998__247__R1_0/}
}

@article{Portmann.2022,
   title={Security in quantum cryptography},
   volume={94},
   ISSN={1539-0756},
   url={http://dx.doi.org/10.1103/RevModPhys.94.025008},
   DOI={10.1103/revmodphys.94.025008},
   number={2},
   journal={Reviews of Modern Physics},
   publisher={American Physical Society (APS)},
   author={Portmann, Christopher and Renner, Renato},
   year={2022},
   month=jun }

@misc{Renner.2006,
      title={Security of Quantum Key Distribution}, 
      author={Renato Renner},
      year={2006},
      eprint={quant-ph/0512258},
      archivePrefix={arXiv},
      primaryClass={quant-ph},
      url={https://arxiv.org/abs/quant-ph/0512258}, 
}

@misc{Arqand.2025,
    author = {Arqand, Amir and Tan, Ernest Y.-Z.},
    title = {Marginal-constrained entropy accumulation theorem},
    note = {arXiv:2502.02563}
}

@misc{Himbeeck.2025,
    author = {Van Himbeeck, Thomas and Brown, Peter},
    title = {A tight and general finite-size security proof
for Quantum Key Distribution},
    note = {to appear}
}

@misc{Watrous.2004,
      title={Notes on super-operator norms induced by Schatten norms}, 
      author={John Watrous},
      year={2004},
      eprint={quant-ph/0411077},
      archivePrefix={arXiv},
      primaryClass={quant-ph},
      url={https://arxiv.org/abs/quant-ph/0411077}, 
}

@article{Beigi.2016,
  title={Hypercontractivity and the logarithmic Sobolev inequality for the completely bounded norm},
  author={Beigi, Salman and King, Christopher},
  journal={Journal of Mathematical Physics},
  volume={57},
  number={1},
  year={2016},
  publisher={AIP Publishing}
}

@inproceedings{Bardet.2022,
  title={Hypercontractivity and logarithmic Sobolev inequality for non-primitive quantum Markov semigroups and estimation of decoherence rates},
  author={Bardet, Ivan and Rouz{\'e}, Cambyse},
  booktitle={Annales Henri Poincar{\'e}},
  volume={23},
  number={11},
  pages={3839--3903},
  year={2022},
  organization={Springer}
}

@misc{Cheng.2024,
      title={Joint State-Channel Decoupling and One-Shot Quantum Coding Theorem}, 
      author={Hao-Chung Cheng and Frédéric Dupuis and Li Gao},
      year={2024},
      eprint={2409.15149},
      archivePrefix={arXiv},
      primaryClass={quant-ph},
      url={https://arxiv.org/abs/2409.15149}, 
}

@article{Gupta.2015,
  author    = {Manish K. Gupta and Mark M. Wilde},
  title     = {Multiplicativity of Completely Bounded p-Norms Implies a Strong Converse for Entanglement-Assisted Capacity},
  journal   = {Communications in Mathematical Physics},
  year      = {2015},
  volume    = {334},
  number    = {2},
  pages     = {867--887},
  doi       = {10.1007/s00220-014-2212-9},
  url       = {https://doi.org/10.1007/s00220-014-2212-9},
  issn      = {1432-0916}
}

@article{Winick.2018,
   title={Reliable numerical key rates for quantum key distribution},
   volume={2},
   ISSN={2521-327X},
   url={http://dx.doi.org/10.22331/q-2018-07-26-77},
   DOI={10.22331/q-2018-07-26-77},
   journal={Quantum},
   publisher={Verein zur Forderung des Open Access Publizierens in den Quantenwissenschaften},
   author={Winick, Adam and Lütkenhaus, Norbert and Coles, Patrick J.},
   year={2018},
   month=jul, pages={77} }

@article{Wilde.2014,
  author    = {Mark M. Wilde and Andreas Winter and Dong Yang},
  title     = {Strong Converse for the Classical Capacity of Entanglement-Breaking and Hadamard Channels via a Sandwiched Rényi Relative Entropy},
  journal   = {Communications in Mathematical Physics},
  year      = {2014},
  volume    = {331},
  number    = {2},
  pages     = {593--622},
  doi       = {10.1007/s00220-014-2122-x},
  url       = {https://doi.org/10.1007/s00220-014-2122-x},
  issn      = {1432-0916}
}

@article{Lennert.2013,
    author = {Müller-Lennert, Martin and Dupuis, Frédéric and Szehr, Oleg and Fehr, Serge and Tomamichel, Marco},
    title = {On quantum Rényi entropies: A new generalization and some properties},
    journal = {Journal of Mathematical Physics},
    volume = {54},
    number = {12},
    pages = {122203},
    year = {2013},
    month = {12},
    issn = {0022-2488},
    doi = {10.1063/1.4838856},
    url = {https://doi.org/10.1063/1.4838856},
    eprint = {https://pubs.aip.org/aip/jmp/article-pdf/doi/10.1063/1.4838856/15705273/122203\_1\_online.pdf},
}

@INPROCEEDINGS{Gao.2023,
  author={Gao, Li and Junge, Marius and LaRacuente, Nicholas},
  booktitle={2018 IEEE International Symposium on Information Theory (ISIT)}, 
  title={Uncertainty Principle for Quantum Channels}, 
  year={2018},
  volume={},
  number={},
  pages={996-1000},
  doi={10.1109/ISIT.2018.8437730}}

@article{Metger.2024,
  author       = {Tony Metger and Omar Fawzi and David Sutter and Renato Renner},
  title        = {Generalised Entropy Accumulation},
  journal      = {Communications in Mathematical Physics},
  volume       = {405},
  number       = {11},
  pages        = {261},
  year         = {2024},
  date         = {2024-10-12},
  issn         = {1432-0916},
  doi          = {10.1007/s00220-024-05121-4},
  url          = {https://doi.org/10.1007/s00220-024-05121-4}
}

@article{Lieb.1978,
  title={Proof of an entropy conjecture of Wehrl},
  author={Lieb, Elliott H},
  journal={Communications in Mathematical Physics},
  volume={62},
  number={1},
  pages={35--41},
  year={1978},
  publisher={Springer}
}

@article{Dembo.1991,
  title={Information theoretic inequalities},
  author={Dembo, Amir and Cover, Thomas M and Thomas, Joy A},
  journal={IEEE Transactions on Information theory},
  volume={37},
  number={6},
  pages={1501--1518},
  year={1991},
  publisher={IEEE}
}

@article{Beigi.2023,
   title={Operator-valued Schatten spaces and quantum entropies},
   volume={113},
   ISSN={1573-0530},
   url={http://dx.doi.org/10.1007/s11005-023-01712-9},
   DOI={10.1007/s11005-023-01712-9},
   number={5},
   journal={Letters in Mathematical Physics},
   publisher={Springer Science and Business Media LLC},
   author={Beigi, Salman and Goodarzi, Milad M.},
   year={2023},
   month=aug }

@article{Arimoto.1977,
  title={Information measures and capacity of order $\alpha$ for discrete memoryless channels},
  author={Arimoto, Suguru},
  journal={Topics in information theory},
  year={1977},
  publisher={The Netherlands}
}

@article{Devetak.2006,
   title={Multiplicativity of Completely Bounded p-Norms Implies a New Additivity Result},
   volume={266},
   ISSN={1432-0916},
   url={http://dx.doi.org/10.1007/s00220-006-0034-0},
   DOI={10.1007/s00220-006-0034-0},
   number={1},
   journal={Communications in Mathematical Physics},
   publisher={Springer Science and Business Media LLC},
   author={Devetak, Igor and Junge, Marius and King, Christoper and Ruskai, Mary Beth},
   year={2006},
   month=may, pages={37–63} }

@article{Bardet.2024,
    title={Entropy decay for Davies semigroups of a one dimensional quantum lattice},
    author={Bardet, Ivan and Capel, {\'A}ngela and Gao, Li and Lucia, Angelo and P{\'e}rez-Garc{\'\i}a, David and Rouz{\'e}, Cambyse},
    journal={Communications in Mathematical Physics},
    volume={405},
    number={2},
    pages={42},
    year={2024},
    publisher={Springer}
}

@article{Metger.2023,
  title={Security of quantum key distribution from generalised entropy accumulation},
  author={Metger, Tony and Renner, Renato},
  journal={Nature Communications},
  volume={14},
  number={1},
  pages={5272},
  year={2023},
  publisher={Nature Publishing Group UK London}
}

@article{Dupuis.2020,
  title={Entropy accumulation},
  author={Dupuis, Frederic and Fawzi, Omar and Renner, Renato},
  journal={Communications in Mathematical Physics},
  volume={379},
  number={3},
  pages={867--913},
  year={2020},
  publisher={Springer}
}

@article{Xu_2020,
   title={Secure quantum key distribution with realistic devices},
   volume={92},
   ISSN={1539-0756},
   url={http://dx.doi.org/10.1103/RevModPhys.92.025002},
   DOI={10.1103/revmodphys.92.025002},
   number={2},
   journal={Reviews of Modern Physics},
   publisher={American Physical Society (APS)},
   author={Xu, Feihu and Ma, Xiongfeng and Zhang, Qiang and Lo, Hoi-Kwong and Pan, Jian-Wei},
   year={2020},
   month=may }

@article{Liao_2017,
   title={Satellite-to-ground quantum key distribution},
   volume={549},
   ISSN={1476-4687},
   url={http://dx.doi.org/10.1038/nature23655},
   DOI={10.1038/nature23655},
   number={7670},
   journal={Nature},
   publisher={Springer Science and Business Media LLC},
   author={Liao, Sheng-Kai and Cai, Wen-Qi and Liu, Wei-Yue and Zhang, Liang and Li, Yang and Ren, Ji-Gang and Yin, Juan and Shen, Qi and Cao, Yuan and Li, Zheng-Ping and Li, Feng-Zhi and Chen, Xia-Wei and Sun, Li-Hua and Jia, Jian-Jun and Wu, Jin-Cai and Jiang, Xiao-Jun and Wang, Jian-Feng and Huang, Yong-Mei and Wang, Qiang and Zhou, Yi-Lin and Deng, Lei and Xi, Tao and Ma, Lu and Hu, Tai and Zhang, Qiang and Chen, Yu-Ao and Liu, Nai-Le and Wang, Xiang-Bin and Zhu, Zhen-Cai and Lu, Chao-Yang and Shu, Rong and Peng, Cheng-Zhi and Wang, Jian-Yu and Pan, Jian-Wei},
   year={2017},
   month=aug, pages={43–47} }

@article{Arnon.2018,
  title={Practical device-independent quantum cryptography via entropy accumulation},
  author={Arnon-Friedman, Rotem and Dupuis, Fr{\'e}d{\'e}ric and Fawzi, Omar and Renner, Renato and Vidick, Thomas},
  journal={Nature communications},
  volume={9},
  number={1},
  pages={459},
  year={2018},
  publisher={Nature Publishing Group UK London}
}

\newpage

\appendix

\section{Some properties of completely bounded norms and a proposition}\label{app:Proof.of.baisc.properties}

In order to compute certain stabilized divergences involving a system $Q$, the system that acts as the `environment' can often be assumed to have the same dimension as that of $Q$. Here we derive a general statement of this kind: in words, the optimization over environment systems $E$ inherent in the cb-norm of some linear map is attained on `environments' $E$ that are isomorphic to the input system $A$ of the map. Since for separable systems this statement is trivial we focus here on the finite dimensional one.
\begin{lemma} \label{lem:finiteEnvironement}
   Let $\Phi:\mathcal{B}(\mathcal{H}_A)\to \mathcal{X}$ be a linear map onto an arbitrary operator space $\mathcal{X}$, with $d_A:=\dim(\mathcal{H}_A)<\infty$. Then, for any $p\ge 1$,
   \begin{align}
\sup_{d\in\mathbb{N}}\|\id_{d}\otimes \Phi\|_{\mathcal{S}_1(\mathbb{C}^d\otimes \mathcal{H}_A)\to \mathcal{S}_p[\mathbb{C}^d,\mathcal{X}]}=\|\id_{d_A}\otimes \Phi\|_{\mathcal{S}_1(\mathbb{C}^{d_A}\otimes \mathcal{H}_A)\to \mathcal{S}_p[\mathbb{C}^{d_A},\mathcal{X}]}
   \end{align}   
\end{lemma}
Note that when $d_A:=\dim(\mathcal{H}_A)=\infty$ the statement is clearly true in the sense that then the environment system $E$ for which the supremum is achieved is isomorphic to system $A$.
The proof of \cref{lem:finiteEnvironement} is a standard consequence of the Schmidt decomposition.

\begin{proof}[Proof of \cref{lem:finiteEnvironement}]
For notational convenience, we denote the norm $\|.\|_{\mathcal{S}_p[\mathbb{C}^d,\mathcal{X}]}$ by $\|.\|_{\#}$ as well as the norm $\|.\|_{\mathcal{S}_1(\mathbb{C}^d\otimes \mathcal{H}_A)\to \mathcal{S}_p[\mathbb{C}^d,\mathcal{X}]}$ by $\|.\|_{1\to \#}$. Clearly $\|\id_d\otimes \Phi\|_{1\to \#}$ is a non-decreasing sequence in $d$, hence it suffices to show that for $d\geq d_A$ it is non-increasing. Fix some $d\geq d_A$, then $\rho\mapsto\|(\id_d\otimes \Phi)(\rho)\|_{\#}$ is convex, hence the maximum is achieved on the extremal operators in $\{\omega\in\mathcal{B}(\mathbb{C}^d\otimes \mathcal{H}_A)|\|\omega\|_1\leq 1\}$, which are just rank-1 operators of the form $\omega=|\psi\rangle\langle\varphi|$: 
\begin{align}
   \sup_{\|\omega\|_1\le 1} \|(\id_d\otimes \Phi)(\omega)\|_{\#} = \sup_{|\varphi\rangle,|\psi\rangle\in\mathbb{C}^d\otimes\mathcal{H}_A, \||\varphi\rangle\|,\,\||\psi\rangle\|\leq 1} \|(\id_d\otimes \Phi)(|\psi\rangle\langle\varphi|)\|_{\#}\,.
\end{align}

Now for any such $|\psi\rangle\in\mathbb{C}^d\otimes\mathcal{H}_A$, by Schmidt decomposition there exists a positive trace-normalized operator $\omega\in\mathcal{S}(\mathcal{H}_A)$ and a local isometry $U:\mathbb{C}^{d_A}\to \mathbb{C}^d$, s.t. 
\begin{align}
    |\psi\rangle=(U\otimes\1)\sum_{i=1}^{d_A}(\1\otimes\sqrt{\omega})(|i\rangle_{\mathbb{C}^{d_A}}\otimes|i\rangle_{\mathcal{H}_A}) \equiv (U\otimes\1)|\sqrt{\omega}\rangle,
\end{align} i.e. $|\psi\rangle$ is a purification of some state $\omega$.  Since $\#$ is invariant under local isometries (cf. Corollary \ref{cor:Pisier.Formula}), it follows that
\begin{align}
    \sup_{\psi,\varphi} \|(\id_d\otimes\Phi)(|\psi\rangle\langle\varphi|)\|_{\#}=\sup_{\omega,\eta\in S(\mathcal{H}_A)}\|(\id_{d_A}\otimes\Phi)(|\sqrt{\omega}\rangle\langle\sqrt{\eta}|)\|_\# \leq \|\id_{d_A}\otimes\Phi\|_{1\to \#},
\end{align} since $\||\sqrt{\omega}\rangle\langle\sqrt{\eta}|\|_1\le\|\omega\|_1\|\eta\|_1\le 1$.
\end{proof}

Next, when considering the norm of a CP map, often one can restrict the optimization to run over over positive-semidefinite inputs, see. e.g. \cite{Watrous.2004} for $q\to p$ and \cite{Devetak.2006} for CB $q\to p$ norms. We generalize some of these statements here to norms of CP maps $\Phi:Q\to RS$ between operator-valued Schatten spaces. 
\begin{lemma}\label{lem:positivesufficiency} 
Let $\Phi:Q\to RS$ be a CP map. Then for $q\leq r\leq s$
\begin{align}
    \|\Phi\|_{cb, (Q:q)\to (R:r,S:s)}= \|\Phi\|^+_{cb, (Q:q)\to (R:r,S:s)},
\end{align} and moreover for $q \geq r,s$
\begin{align}
    \|\Phi\|_{cb,( Q:q)\to (R:r,S:s)}= \|\Phi\|^+_{(Q:q)\to (R:r,S:s)}.
\end{align} 
\end{lemma}
These are generalizations of \cite[Theorem 12 and 13 respectively]{Devetak.2006}, given our \cref{cor:variationalFormula}. Before giving their proof  we require the following Lemma, which is a generalization of \cite[Lemma 9]{Devetak.2006}.
\begin{proposition} \label{prop:ContractionInequality}
Let $r\leq t\leq s$ and $X\in B(\mathcal{H}_{123})$ be a contraction, then
\begin{align}
    \|C^*XD\|_{(r,t,s)}\leq \|C^*C\|_{(r,t,s)}^{\frac{1}{2}}\|D^*D\|_{(r,t,s)}^{\frac{1}{2}},
\end{align} where for notational simplicity we dropped the labeling in the notation, i.e. wrote $\|\cdot\|_{(r,t,s)}\equiv \|\cdot\|_{(1:r,2:t,3:s)}$, etc. 
\end{proposition}
\begin{proof}
By assumptions there exists, due to Lemma \ref{thm:VariationalSpOperatorspace}, $A,B\in S_{2x}(\mathcal{H}_1)$, with $\frac{1}{x}=\frac{1}{r}-\frac{1}{t}$ s.t. $\|A\|_{2x}=\|B\|_{2x}=1$, $A,B\geq 0$, and $Y,Z\geq 0$ such that
\begin{align}
    C^*C=(A_1\otimes\1_{23})Y(A_1\otimes\1_{23}) \quad \& \quad \|C^*C\|_{(r,t,s)}=\|Y\|_{(t,t,s)}, \\
    D^*D=(B_1\otimes\1_{23})Z(B_1\otimes\1_{23}) \quad \& \quad \|D^*D\|_{(r,t,s)}=\|Z\|_{(t,t,s)}. \\
\end{align} 
There further exist an isometries $V, W$ s.t.
\begin{align}
    C=VY^{\frac{1}{2}}(A\otimes \1) \quad     D=WZ^{\frac{1}{2}}(B\otimes \1),
\end{align} and hence $C^*XD=(A\otimes\1)Y^{\frac{1}{2}}V^*XWZ^{\frac{1}{2}}(B\otimes\1)$. So Lemma \ref{thm:VariationalSpOperatorspace} together with \cite{Devetak.2006}[Lemma 9] yield
\begin{align}
    \|C^*XD\|_{(r,t,s)}\leq \|Y^{\frac{1}{2}}V^*XWZ^{\frac{1}{2}}\|_{(t,t,s)}\leq \|Y\|^{\frac{1}{2}}_{(t,t,s)}\|Z\|^{\frac{1}{2}}_{(t,t,s)} = \|C^*C\|^{\frac{1}{2}}_{(r,t,s)}\|D^*D\|^{\frac{1}{2}}_{(r,t,s)},
\end{align}
which is what we wanted to prove.
\end{proof}

We may now give the proof of \cref{lem:positivesufficiency}.
\begin{proof}
We first proof the first part. 
Fix some environment $E$ and let $Q\in B(EQ)$. Let $Q=U|Q|^{\frac{1}{2}}|Q|^{\frac{1}{2}}$ be its polar decomposition. Then
\begin{align}
   0 < \begin{pmatrix}
    U|Q|^{\frac{1}{2}} \\
    |Q|^{\frac{1}{2}}
    \end{pmatrix} (|Q|^{\frac{1}{2}}U^* \quad |Q|^{\frac{1}{2}}) = \begin{pmatrix}
        U|Q|U^* \quad Q \\
        Q^* \quad |Q|
    \end{pmatrix}.
\end{align}
Now since $\Phi$ is CP, $\id_2\otimes\id_E\otimes\Phi$ is positive and hence so is
\begin{align}
    \begin{pmatrix}
        (\id_E\otimes\Phi)(U|Q|U^*) \quad (\id_E\otimes\Phi)(Q) \\
        (\id_E\otimes\Phi)(Q^*) \quad (\id_E\otimes\Phi)(|Q|)
    \end{pmatrix} >0.
\end{align}
Now it is well known \cite{Book.Hiai.2014} that this block matrix being positive is equivalent to 
\begin{align}
\label{equ:ineq1}
    (\id_E\otimes\Phi)(Q) = (\id_E\otimes\Phi)(U|Q|U^*)^{\frac{1}{2}}X((\id_E\otimes\Phi)(|Q|))^{\frac{1}{2}}
\end{align}
for some contraction $X$. Now directly \eqref{equ:ineq1} and \cref{prop:ContractionInequality} yield the desired result, since $q\leq r\leq s$ and
\begin{align}
  \|(\id_E\otimes\Phi)(Q) \|_{(E:q,R:r,S:s)} &=  \|(\id_E\otimes\Phi)(U|Q|U^*)^{\frac{1}{2}}X((\id_E\otimes\Phi)(|Q|))^{\frac{1}{2}}\|_{(E:q,R:r,S:s)} \\ &\leq \|(\id_E\otimes\Phi)(U|Q|U^*)\|^{\frac{1}{2}}_{(E:q,R:r,S:s)}\|(\id_E\otimes\Phi)(|Q|)\|^{\frac{1}{2}}_{(E:q,R:r,S:s)} \\ 
  &\leq \|\id_E\otimes\Phi\|^+_{(E:q,Q:q)\to (E:q,R:r,S:s)} \|U|Q|U^*\|^{\frac{1}{2}}_{(E:q,Q:q)}\||Q|\|^{\frac{1}{2}}_{(E:q,Q:q)}
  \\ &=  \|\id_E\otimes\Phi\|^+_{(E:q,Q:q)\to (E:q,R:r,S:s)}\|Q\|_{q},
\end{align} where in the last equality we used unitary invariance in of the Schatten-$p-$norm. Taking the supremum over $E$ gives the result. \\
To prove the second part we use the fact that the SWAP operator is a complete contraction, see \eqref{equ:complete.contraction}, hence since $q\geq r,s$
\begin{align}
    \|\Phi\|_{cb,Q:q\to (R:r,S:s)} &= \sup_E\sup_{X_{EQ}}\frac{\|(\id_E\otimes\Phi)(X_{EQ})\|_{(E:q,R:r,S:s)}}{\|X_{EQ}\|_{(E:q,Q:q)}} \\ &= \sup_E\sup_{X_{EQ}}\frac{\|(\id_E\otimes\Phi)(X_{EQ})\|_{(E:q,R:r,S:s)}}{\|X_{QE}\|_{q}} \\&\leq \sup_E\sup_{X_{QE}}\frac{\|(\Phi\otimes \id_E)(X_{QE})\|_{(R:r,S:s,E:q)}}{\|X_{QE}\|_{q}} \\ &\leq \|\Phi\|^+_{Q:q\to (R:r,S:s)},
\end{align} where the first inequality are two applications of the SWAP operator and the last inequality is \cref{lem:reductionLemma}.
\end{proof}

\section{Additivity on reduced state space}\label{app:Additivity.on.reduced.operator.spaces}


Now we are in a position to prove \cref{thm:Restricted.multiplicativity} starting from \cref{thm:mainchainrule} using an argument from \cite{Himbeeck.2025}. There it was used to prove a version of \cref{thm:IIDreduction} under $n$-fold tensor products of channels.
\begin{proof}[Proof of \cref{thm:Restricted.multiplicativity}]
To fix $n$ linear constraints fix some CPTP channels $\{\mathcal{N}_i:Q_i\to Q^\prime_i\}_{i=1}^n$ and states $\{\tau_i\in\mathcal{D}(\cH_{Q^\prime_i})\}_{i=1}^n$. Recall the definitions of $Q^n:=Q_1...Q_n$, $\mathcal{D}_r(\cH_{Q})$ and $\mathcal{D}^E_r(\cH_{Q})$.
WLOG assume that none of the $\Phi_i\neq 0,$ else the statement trivially holds.

To do this we first recall, that due to \cref{lem:finiteEnvironement} and its proof it holds that 
\begin{align}
   \|\Phi\|_{r,cb,Q:1\to(R:1,S:p)} &= \sup_{E}\sup_{\rho_{EQ}\in\mathcal{D}^E_r(\cH_Q)}\|(\id_E\otimes\Phi)(\rho_{EQ})\|_{(E:1,R:1,S:p)} \\ &= \sup_{\rho\in\mathcal{D}_r(\cH_Q)}\|(\id_{\tilde{Q}}\otimes\Phi)(|\sqrt{\rho}\rangle\langle\sqrt{\rho}|_{\tilde{Q}Q})\|_{(\tilde{Q}:1,R:1,S:p)} \\
   &=:\sup_{\rho_\in\mathcal{D}_r(\cH_{Q})}g_p^{\Phi}(\rho_Q) \\ &=\sup\{g_p^\Phi(\rho_Q)| \rho_Q\geq 0, \mathcal{N}(\rho_Q)=\tau\},
\end{align} where $\tilde{Q}$ is isomorphic to $Q$ and $|\sqrt{\rho}\rangle_{\tilde{Q}Q}=(\1_{\tilde{Q}}\otimes\sqrt{\rho}_Q)\sum_{i}|ii\rangle_{\tilde{Q} Q}$ is a (canonical) purification of $\rho_Q\in\mathcal{D}_r(\cH_{Q})$ into $\tilde{Q} Q$. This is because the extremal points of the linear constrained $\mathcal{D}_r^E(\cH_{Q}):=\{\rho\in\mathcal{D}(\cH_{EQ})|(\mathcal{N}\circ \tr_E)(\rho_{EQ})=\tau\}$ are nothing but the set of purifications of all $\rho\in\mathcal{D}_r(\cH_{Q})$, by construction. To get the last line we used that the linear constraint already enforces $\Tr[\rho]=\Tr[\mathcal{N}(\rho)]=\Tr[\tau]=1$, since $\mathcal{N}$ is TP.

Hence the statement of the theorem is equivalent to
\begin{align}\label{equ:simplification.C}
\sup_{\rho\in\mathcal{D}_r(\cH_{Q^n})}g_p^{\otimes_{i=1}^n\Phi_i}(\rho) = \prod_{i=1}^n \sup_{\rho_i\in\mathcal{D}_r(\cH_{Q_i})} g^{\Phi_i}_p(\rho_i).   
\end{align}
Since for operators $\rho_i\in\mathcal{D}_r(\cH_{Q_i})$ it clearly holds that $\otimes_{i=1}^n\rho_i\in\mathcal{D}_r(\cH_{Q^n})$ it actually suffices to prove that the LHS in \eqref{equ:simplification.C} is upper bounded by the RHS.
To do so the authors in \cite{Himbeeck.2025} have shown that in the finite dimensional setting, the above convex optimization problem is equal to its dual, i.e.

\begin{align}\label{equ:Strong.Duality}
    \sup_{\rho\in\Pos(\mathcal{H}_Q)}\{g^\Phi_p(\rho)|\mathcal{N}(\rho)=\tau\} = \inf_{\Sigma\in\Pos(\mathcal{H}_{Q^\prime})}\{\Tr[\Sigma\tau]| g^\Phi_p(\rho)\leq \Tr[\Sigma\mathcal{N}(\rho)] \forall \rho\in\Pos(\mathcal{H}_Q)\}.
\end{align}
This follows by showing that there exists a dual feasible $\Sigma\in\Pos(\mathcal{H}_{Q^\prime})$, i.e. one that satisfies
\begin{align}
    g_p^\Phi(\rho)<\Tr[\Sigma \mathcal{N}(\rho)] \ \forall \rho\in\Pos(\mathcal{H}_Q).
\end{align}

For the convenience of the reader we will give a simple proof. Note first that both sides are positive homogeneous in $\rho$, hence it suffices to show it for all $\rho\in\mathcal{D}(\cH_{Q})$. Let $\Sigma=C\1_{Q^\prime}$, then the RHS becomes $\Tr[\Sigma \mathcal{N}(\rho)]=C$, since $\mathcal{N}$ is CPTP and this inequality holds strictly, by definition for $C =\|\Phi\|_{cb,Q:1\to(R:1,S:p)} + 1<\infty$ by assumption. 
Now that we have strong duality, let $\Sigma_i$ be a feasible point of the dual problem, i.e.
for any $0\neq \rho\in\Pos(\mathcal{H}_{Q_i})$, $0<g_p^{\Phi_i}(\rho)\leq \Tr[\mathcal{N}^*_i(\Sigma_i)\rho]$, since $\Phi_i\neq 0$.
It follows that $\mathcal{N}^*(\Sigma_i) > 0$. Define the hence  strictly positive $V_i:=\sqrt{\mathcal{N}^*_i(\Sigma_i)}$ and with it the CP map $\Phi_i^\prime(\cdot):=\Phi_i(V_i^{-1}\cdot V_i^{-1})$. 
Now we see that
\begin{align}
g_p^{\Phi_i}(\rho) &\leq \Tr[\mathcal{N}^*_i(\Sigma)\rho] \quad &\forall \rho\in\Pos(\mathcal{H}_{Q_i}), \\ 
\Leftrightarrow  g_p^{\Phi^\prime_i}(\rho^\prime) &\leq  \Tr[\rho_i^\prime] \quad &\forall \rho^\prime \in \Pos(\mathcal{H}_{Q_i}), \\
\Leftrightarrow g_p^{\Phi^\prime_i}(\rho^\prime) &\leq 1 &\forall \rho^\prime\in\mathcal{D}(\cH_{Q_i}), \\
{\Rightarrow} g_p^{\otimes_{i=1}^n\Phi^\prime_i}(\rho_n^\prime) &\leq 1 \quad &\forall \rho_n^\prime\in \mathcal{D}(\cH_{Q^n}), \\ 
\Leftrightarrow g_p^{\otimes_{i=1}^n\Phi^\prime_i}(\rho_n^\prime) &\leq \Tr[\rho_n^\prime] \quad &\forall \rho_n^\prime\in\Pos(\mathcal{H}_{Q^n}), \\
\Leftrightarrow g_p^{\otimes_{i=1}^n\Phi_i}(\rho_n) &\leq \Tr[\otimes^n_{i=1}\mathcal{N}^*_i(\Sigma_i)\rho_n] = \Tr[(\otimes_{i=1}^n\Sigma_i)(\otimes_{i=1}^n\mathcal{N}_i)(\rho_n)] 
\quad &\forall \rho_n\in\Pos(\mathcal{H}_{Q^n}).
\end{align} where $\rho^\prime:=V_i\rho V_i$ and $\rho_n=(\bigotimes_{i=1}^nV_i^{-1})\rho^\prime_n(\bigotimes_{i=1}^nV_i^{-1})$\,, 
where the implication above followed from \cref{thm:mainchainrule}. 
So we have shown that  $\otimes_{i=1}^n \Sigma_i$ is dual feasible. In addition, the objective value for $\otimes_{i=1}^n \Sigma_i$ is $\prod_{i=1}^n \Tr[\Sigma_i \tau_i]$.
Taking the infimum over all feasible $\Sigma_i$ for $i\in[n]$ yields the claim due to strong duality \eqref{equ:Strong.Duality}.

\end{proof}

\section{Uniform continuity for $f$-weighted Renyi entropies}

\begin{lemma}
    \label{lem:continutity} 
    Consider systems $AE$ with finite-dimensional Hilbert spaces, a classical system $X$ with basis elements labeled by $\mathbb{X}$ and a function $f:\mathbb{X}\to \mathbb{R}$. Define $\eta_0 = |A| (2^{\max_x f(x)} + 2^{-\min_x f(x)}) + 1$. For $\alpha \in (1,1+\frac{1}{\log \eta_0})$, we have for all states $\rho \in \cD(\cH_{EXA})$
    \begin{align}
    H(A|XE)_{\rho} - \mathbb E_{x \sim \rho_X}[f(X)] - (\alpha-1) (\log \eta_0)^2 \leq H^{\uparrow,f}_\alpha(A|XE)_{\rho} \leq H(A|XE)_{\rho} - \mathbb E_{x \sim \rho_X}[f(X)]\,.
    \end{align}
\end{lemma}
\begin{proof}
    We can get this result from the uniform continuity of Rényi divergences which are defined for $\alpha\in (0,1)\cup (1,\infty)$ as 
    \begin{align}
        D_\alpha(\rho \| \sigma) = \frac{\alpha}{1-\alpha} \log \|\sigma^{\tfrac{1-\alpha}{2\alpha}} \rho \sigma^{\tfrac{1-\alpha}{2\alpha}}\|_{\alpha}\\
        D'_\alpha(\rho \| \sigma) = \frac{1}{\alpha-1} \log \Tr[\rho^{\alpha} \sigma^{1-\alpha}]
    \end{align}
    and for $\alpha=0$ as $D'_0(\rho \| \sigma) = \lim_{\alpha\to 0}D'_\alpha(\rho \| \sigma)$.     We note that \[
    H^{\uparrow,f}_\alpha(A|XE)_{\rho} = - \min_{\sigma\in \cD(\cH_{XE})} D_\alpha(\rho_{AXE} \| 2^{-f_X}\cdot \sigma_{XE}) \geq - D_\alpha(\rho_{AXE} \| 2^{-f_X}\cdot \rho_{XE}) \] 
    and  $H(A|XE)_{\rho} - \mathbb E[f(X)] = - D(\rho_{AXE}\|2^{-f_X}\cdot \rho_{XE}) = - \min_{\sigma \in \cD(\cH_{XE})} D_{\alpha}(\rho_{AXE}\|2^{-f_X}\cdot \rho_{XE})$.
    
    In \cite[Lemma B.8 and Eq. (82)]{Dupuis.2020}  it is proven that, for all $\alpha\in (1,1+\log(\eta))$, 
    \begin{align}
    D(\rho_{AXE} \| 2^{-f_X}\rho_{XE}) \leq D_\alpha(\rho_{AXE} \| 2^{-f_X}\rho_{XE}) \leq D(\rho_{AXE} \| 2^{-f_X}\rho_{XE}) + (\alpha-1) (\log \eta)^2
    \end{align}
    with $\eta = 2^{D'_2(\rho_{AXE} \| 2^{-f_X} \rho_{XE})} + 2^{-D_0(\rho_{AXE} \| 2^{-f_X} \rho_{XE})} + 1$. 
    
 This yields the desired continuity statement, provided we can bound $\eta$ in a state independent way. For this, observe that for all $\alpha \in (1,\infty)$, 
    \begin{align}
        D'_\alpha(\rho_{AXE}\| 2^{-f_{X}}\rho_{XE}) 
            &= \frac{1}{\alpha-1} \log \Tr[\rho_{AXE}^{\alpha} \rho_{XE}^{1-\alpha} 2^{(\alpha-1)f_{X}} ]\\
            &\leq \frac{1}{\alpha-1} \log( \Tr[\rho_{AXE}^{\alpha} \rho_{XE}^{1-\alpha} ] 2^{(\alpha - 1) \max_{x} f(x)}) \\
            &\leq \log |A| + \max_x f(x)\,.
    \end{align}
    Similarly, $-D'_\alpha(\rho_{AXE} \| 2^{-f_X}\rho_{XE}) \leq \log \tr[2^{-f_X} \rho_{XE}] \leq \log |A| - \min_x f(x)$. This implies that 
    \begin{align}
        \log \eta \leq \log (|A|2^{\max_x |f(x)|} + |A|2^{-\min_x f(x)}+1).
    \end{align}
\end{proof}

\end{document}